\documentclass[10pt]{wlscirep}
\usepackage[utf8]{inputenc}
\usepackage[T1]{fontenc}
\usepackage{graphicx}
\usepackage{amssymb}
\usepackage{epstopdf}
\usepackage{graphicx}
\usepackage{amssymb}
\usepackage{epstopdf}
\usepackage{dsfont}
\usepackage{xcolor}
\usepackage{amsmath}
\usepackage{caption} 
\usepackage{subcaption}
\usepackage{tikz}
\usepackage{comment}
\usepackage{minted}
\usepackage[dvipsnames]{xcolor}
\usepackage{listings}
\usepackage{verbatim}
\usepackage{booktabs}
\usepackage{adjustbox}

\lstdefinelanguage{Solidity}{
  keywords=[1]{contract, function, uint, uint256, require, returns, public, private, view, pure, payable, if, else, for, while, emit, event, address, mapping,new,await},
  keywordstyle=[1]\color{blue}\bfseries,
  keywords=[2]{pragma, import, const, connect},
  keywordstyle=[2]\color{red}\bfseries,
  keywords=[3]{true, false},
  keywordstyle=[3]\color{purple},
  keywords=[3]{abi, bytecode},
  keywordstyle=[3]\color{ForestGreen},
  sensitive=true,
  comment=[l]{//},
  morecomment=[s]{/*}{*/},
  commentstyle=\color{gray},
  stringstyle=\color{teal},
  morestring=[b]",
  morestring=[b]'
}

\def\ba#1#2\ea{\begin{align}\label{#1}#2\end{align}}
\title{Price manipulation schemes of new crypto-tokens in decentralized exchanges}

\author[1,2]{Manuel Naviglio}
\author[3]{Francesco Tarantelli}
\author[1,3]{Fabrizio Lillo}
\affil[1]{Scuola Normale Superiore, Pisa, Italy}
\affil[2]{INFN Sezione di Pisa, Largo Pontecorvo 3, I-56127 Pisa, Italy}
\affil[3]{Dipartimento di Matematica, Universit\`a di Bologna, Bologna, Italy}

\affil[ ]{\texttt{manuel.naviglio@sns.it}, \texttt{francesco.tarantell3@unibo.it}, \texttt{fabrizio.lillo@sns.it}}

\begin{abstract}
Blockchain technology has revolutionized financial markets by enabling decentralized exchanges (DEXs) that operate without intermediaries. Uniswap V2, a leading DEX, facilitates the rapid creation and trading of new tokens, which offer high return potential but exposing investors to significant risks. In this work, we analyze the financial impact of newly created tokens, assessing their market dynamics, profitability and liquidity manipulations. Our findings reveal that a significant portion of market liquidity is trapped in honeypots, reducing market efficiency and misleading investors. Applying a simple buy-and-hold strategy, we are able to uncover some major risks associated with investing in newly created tokens, including the widespread presence of rug pulls and sandwich attacks. We extract the optimal sandwich amount, revealing that their proliferation in new tokens stems from higher profitability in low-liquidity pools. Furthermore, we analyze the fundamental differences between token price evolution in swap time and physical time. Using clustering techniques, we highlight these differences and identify typical patterns of honeypot and sellable tokens. Our study provides insights into the risks and financial dynamics of decentralized markets and their challenges for investors.\\$ $ 
\end{abstract}
\begin{document}

\flushbottom
\maketitle
%
%
\thispagestyle{empty}


\section{Introduction}
The advent of blockchain technology has revolutionized financial markets, enabling decentralized platforms that operate independently of centralized intermediaries. Among these innovations, Decentralized Exchanges (DEXs) have emerged as key players in the cryptocurrency ecosystem. Unlike centralized exchanges, DEXs rely on decentralized protocols implemented via smart contracts deployed on blockchains, allowing trustless, peer-to-peer trading without the need for custodial services. This framework has led to the proliferation of cryptocurrency tokens and trading strategies, driven by the innovative mechanism of Automated Market Makers (AMMs)\cite{AMMeco}, which replaces the traditional order book model with a liquidity pool-based system.
One notable example is Uniswap V2 \cite{uni2core, analysisUniswap}, a widely-used DEX protocol that supports the creation of liquidity pools for token pairs, typically involving {Wrapped} Ethereum ({W}ETH). On average, $\sim 15$ new tokens\cite{fallcrypto} paired with Ethereum are introduced hourly on Uniswap V2 between October {2}, 2024, and December 2, 2024. Many of these tokens initially exhibit extremely high returns, creating an illusion of huge profits and attracting inexperienced investors \cite{pump-dump, pricemanipulation}. However, beneath this apparent simplicity and profitability lie significant risks and vulnerabilities that can lead to partial or total loss of funds \cite{Momtaz2018InitialCO, ICOrisk, sociotec}. Indeed, the decentralized nature of DEXs introduces a set of security challenges, which have been extensively analyzed in the context of computer security and formal verification methods~\cite{Zhou2022SoKDF}. These studies highlight that these coins and many investments hide important vulnerabilities, exposing users to risks that require careful evaluation \cite{aspembitova2021behavioral}. {Among these, the so-called \textit{honeypot} contracts---first formally defined in~\cite{TorresHoney}---are deceptive smart contracts that mislead users by appearing profitable, encouraging deposits that cannot later be withdrawn due to hidden restrictions, effectively trapping the funds. For instance, a honeypot may include concealed conditions or access controls that appear harmless on interfaces such as Etherscan, but are enforced during execution. These mechanisms often rely on hidden \textit{backdoors}, such as black- or whitelists, sell taxes, or non-renounced ownership~\cite{honeypot_url,Sun2024SoKCA}.}

{\bf Contribution of the work.} To our knowledge, most existing studies on the vulnerabilities and risks associated with decentralized exchanges, including phenomena such as honeypots and backdoors, have primarily approached the topic from the perspective of computer security and formal verification\cite{defi_securities}. Although these works have significantly advanced methods to detect malicious contracts and identify vulnerabilities in smart contracts\cite{tokenaware, DApps}, they often lack a quantitative and financial perspective \cite{cryptofinancial}. Specifically, numerous methodologies have been proposed to identify honeypots or detect vulnerabilities, leveraging static and dynamic analysis techniques, formal methods, and auxiliary service data~\cite{Qin2021QuantifyingBE,Zhou2020HighFrequencyTO,HoneypotWeb1,HoneypotWeb2,TorresHoney,honeypot_url,Sun2024SoKCA}. However, there has been little to no contribution focused on understanding the broader financial impact of these phenomena, particularly in relation to the scale and significance of the proliferation of new tokens created within the decentralized finance ecosystem \cite{HU2021102462}. In this work, we aim to bridge this gap by performing a quantitative and financial analysis of the creation and proliferation of new tokens on Uniswap V2. \\To achieve this, we systematically collect and analyze data on newly created tokens on Uniswap V2, aiming to establish the true financial scale of the phenomenon and assess the actual value of these tokens. Our findings reveal that a significant portion of market liquidity is trapped in honeypots, effectively removing liquidity from the market and making it less secure and efficient. The seemingly high returns observed for newly created tokens may give the impression that the market is a gold mine, yet honeypots absorb a substantial portion of these profits. Additionally, applying a simple buy-and-hold strategy highlights further critical issues. First, it reveals the widespread presence of tokens designed for rug pulls, a fraudulent practice where developers remove liquidity from the pool, making the token unsellable and leaving investors with worthless assets. Some of these can be identified using a measure introduced in this paper, i.e. Net Traded Value (NTV), which measures the value of new tokens and quantifies the percentage trapped in honeypots. Furthermore, we observe a widespread presence of sandwich attacks targeting newly created tokens. By computing the optimal sandwich attack size, we show that one of the primary reasons for their prevalence in our data is that sandwich attacks become increasingly profitable in low-liquidity pools, such as those associated with newly created tokens. Finally, our strategy reveals fundamental differences when analyzing token price evolution in swap time versus physical time, due to the block-based structure of transactions. While price behavior in swap time may suggest that these tokens offer frequent opportunities for profit, this perspective is significantly modified when viewed in physical time. To study this effect in detail, we apply clustering techniques to analyze price trajectories both in swap time and in physical time, demonstrating how these different temporal frameworks lead to different interpretations of token performance and market dynamics.

{\bf The organization of the paper.} The paper is organized as follows. Section~\ref{sec:Data} outlines the dataset used in our analysis, detailing the classification of newly created tokens as honeypots or sellable coins. In Section~\ref{sec:Properties}, we quantify the scale of new token creation on Uniswap V2, highlighting how a significant portion of market liquidity remains trapped in honeypots. Section~\ref{sec:B&H} examines the profitability of a buy-and-hold strategy in early-stage tokens. In Section~\ref{sec:Limits}, we discuss the criticalities related to this strategies. Introducing a measure of NTV, we quantify the real value of newly created tokens and assess liquidity risks, including rug pulls and major withdrawals by token creators. We also show the crucial role sanwich attacks have in our strategy and derive the optimal sandwich attack size, explaining their prevalence in low-liquidity pools. Lastly, we analyze price trajectory patterns using clustering in both \textit{swap time} and \textit{physical time}, identifying key behaviors of honeypots. Conclusions are presented in Section~\ref{sec:Conclusions}.


\section{Dataset and Token Classification}\label{sec:Data}

To conduct our analysis, we collected on-chain data from the Ethereum blockchain, focusing on newly created tokens and their trading activity on DEX. The dataset consists of 17,194 tokens launched between October {2}, 2024, and December 2, 2024. Data collection was carried out by connecting to an Ethereum node using an RPC provider such as Infura or Alchemy via Web3. Newly created pairs were identified by monitoring blockchain events and tracking the creation of token pairs, recording their contract addresses along with the associated tokens (\textit{token\_x}, \textit{token\_y}). This allowed us to follow the introduction of new tokens into the ecosystem. We then collected corresponding swap events to capture the amounts swapped in and out, providing insights into price dynamics and trading activity. Additionally, we extracted liquidity-related data by examining mint and burn events, recording the amounts of liquidity tokens minted and burned to assess liquidity provisioning over time. This dataset serves as the foundation for analyzing token behavior, price evolution, and liquidity trends of early-stage token dynamics in DEX.

{We neglect the new tokens created on Uniswap V3 because only about 1,107 were created within our time window, which is an order of magnitude lower than on Uniswap V2. We also disregard possible interactions with other V3 pools involving the same new tokens created on Uniswap V2, since they amount to approximately 273 tokens-representing a very small fraction. Moreover, most of these tokens are associated with low-liquidity V3 pools.}

{\bf Token Classification: sellable vs honeypot.} As discussed in the Introduction, not all tokens remain \textit{sellable} throughout their lifecycle. Many evolve into \textit{honeypot} contracts, effectively preventing traders from selling and trapping liquidity. To distinguish between these two categories, we implemented a multi-step classification approach.

We performed a time-based analysis, observing each token’s entire trading history to detect any transition into honeypot behavior. We leveraged external security detection services such as \href{https://honeypot.is}{Honeypot.is}, which specializes in identifying honeypots, and \href{https://gopluslabs.io/}{GoPlus Labs}, which provides security scoring for tokens. These services allowed us to systematically verify whether a token exhibited selling restrictions at any point in its lifecycle. Based on this analysis, we categorized tokens at the end of their observed lifetime. Thus, we labeled as honeypots those coins that were either initially classified as such or became honeypots later in their lifecycle. However, our analysis indicates that the majority of honeypots transition into this state within the first few swaps. It is important to note that GoPlus and Honeypot.is use different methods to identify honeypots. The former analyzes the smart contract directly, while the latter simulates sell transactions to determine whether a token can be liquidated. {GoPlus primarily performs static code inspection and applies heuristic blacklists to detect suspicious or malicious functions within the contract, while Honeypot.is adopts a dynamic approach by simulating real sell transactions to verify whether liquidation is actually possible. The two methods are therefore complementary: static analysis may detect inactive or hidden traps in the code that never trigger on-chain, whereas dynamic simulation can capture behavioral restrictions that cannot be inferred from bytecode alone.} In our study, we classify a token as a honeypot if at least one of these two providers flags it as such. 

Tokens that never showed honeypot behavior were classified as \textit{sellable}, comprising approximately 2,080 tokens (\(\simeq 12\%\) of the total). In contrast, tokens that showed locked or blocked sales at any stage were classified as \textit{honeypots}, accounting for around 15,114 tokens (\(\simeq 88\%\) of the total).
{We find that \href{https://honeypot.is}{Honeypot.is} detects 91.6\% of them, while \href{https://gopluslabs.io/}{GoPlus Labs} detects 81.6\%. This shows a substantial agreement between the two providers and shows the robustness of our choice of considering their union.}

\section{Early-Stage Properties of Newly Created Tokens}\label{sec:Properties}
In this section, we analyze the early characteristics of newly created tokens on Uniswap V2, including their creation frequency, classification as sellable tokens or honeypots, and average lifespan. Honeypots significantly distort market dynamics by inflating trading volumes while restricting liquidity. By preventing selling, they create an illusion of high demand and price appreciation, misleading traders and absorbing substantial market volume. Understanding how much liquidity remains locked in these contracts provides insights into the true market potential of sellable tokens and the challenges faced by legitimate projects. A new token lifecycle begins with its creation on the Ethereum blockchain, followed by the establishment of a liquidity pool on Uniswap V2, where the owner pairs the token with {W}ETH to enable trading. Initial trades occur almost immediately \cite{initialcoin}, often within the first block, driven by automated bots and, in some cases, the creator simulating demand. This early activity frequently causes a sharp price surge due to low liquidity. However, this surge is often short-lived, as subsequent selling pressure, sandwich attacks, and liquidity rug-pulls emerge. Analyzing these early-stage price dynamics, which is precisely what we do in this paper, helps distinguish between tokens with genuine growth potential and those exhibiting market manipulation \cite{manipulation}.

The first question to consider when understanding the impact of such coins concerns how many of them are being created. 

\begin{figure}[!htb]
\centering
\begin{subfigure}{0.46\textwidth}
  \centering
  \includegraphics[width=\textwidth]{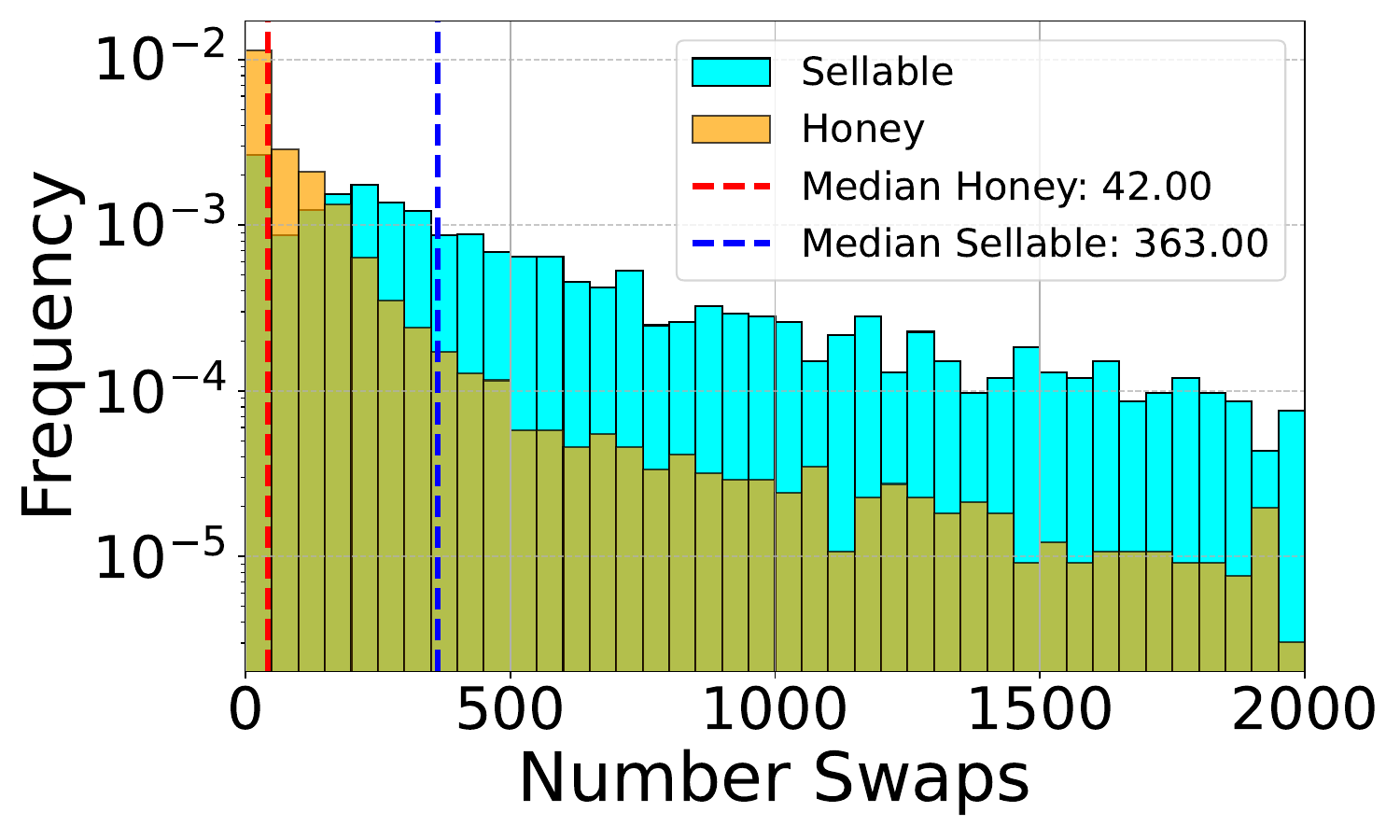}
  \caption{Number of swap distribution.}
  \label{fig:Num_Dist}
\end{subfigure}
\hfill
\begin{subfigure}{0.46\textwidth}
\centering
\includegraphics[width=\linewidth]{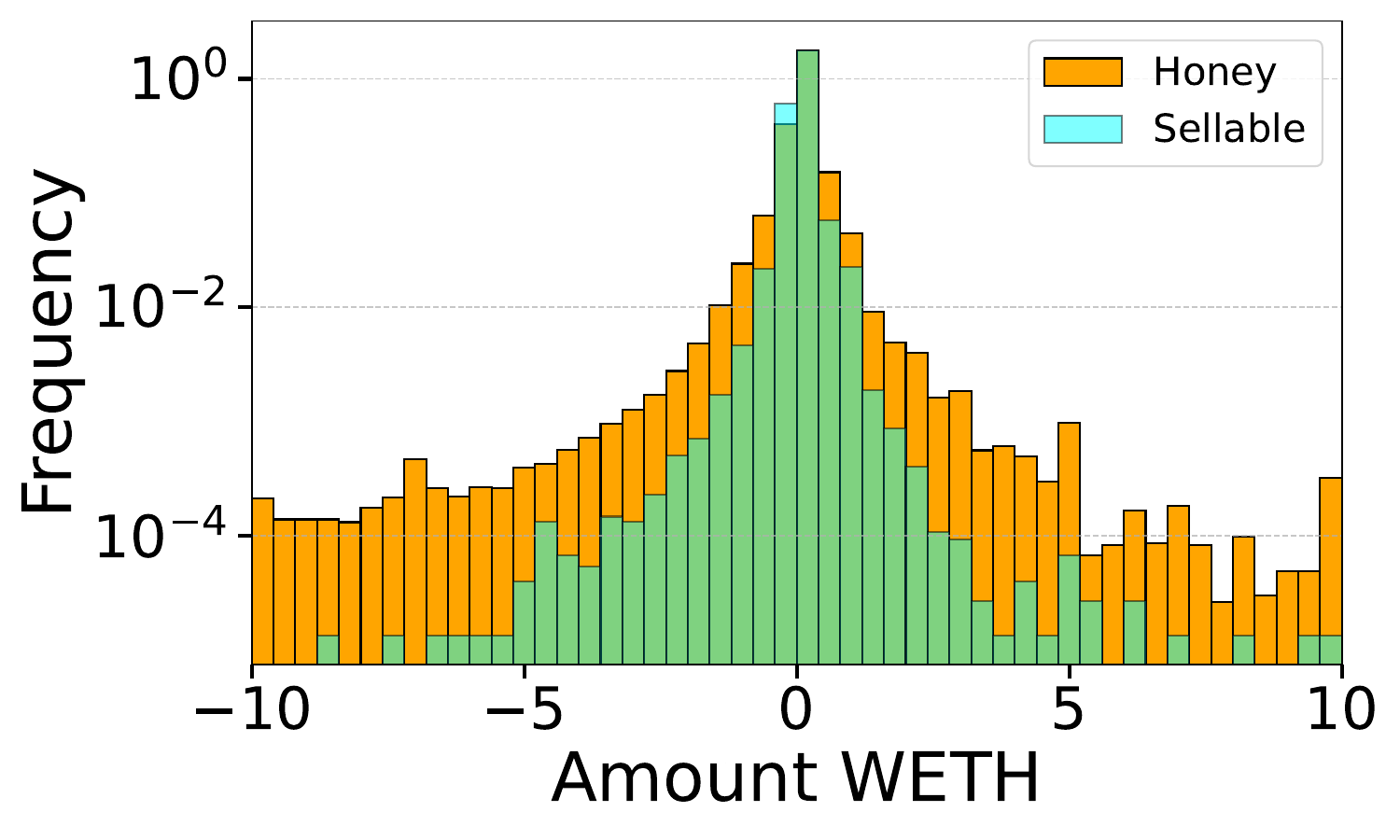}
  \caption{{W}ETH amount distribution of all the swaps.}
  \label{fig:AmountETHDist}
\end{subfigure}
\caption{{Normalized distributions}}
\label{fig:corr_am}
\end{figure}

Remarkably, the number of tokens created daily is on average around 200, highlighting the sustained activity in token creation (see Appendix \ref{app_num_coins}). This trend underscores the importance of Uniswap V2 as a platform to launch and experiment with new tokens. The orange part of the histogram represents the portion of honeypots among these newly created coins. The presence of a substantial fraction of honeypots within this continuous flow of token creation suggests that a significant portion of new projects are designed with malicious intent, aiming to trap liquidity rather than fostering genuine market participation. This dynamic not only skews the perception of overall market activity, but also presents a major challenge for traders and investors, who must navigate an environment where a large share of tokens are inherently unsellable. As a direct consequence, a considerable portion of the liquidity that circulates in the market is effectively locked within honeypot contracts, rendering it inaccessible to traders. This reduces the actual capital available for legitimate trading and investment, further distorting market efficiency. Thus, while the high daily token creation rate may signal strong market engagement, it also reflects an ongoing risk where a substantial amount of liquidity is continuously funneled into fraudulent contracts rather than supporting genuine projects. The peak of the distribution of the tokens number created within one-hour bins occurs around 15. However, the overall distribution is largely dominated by honeypots, which are created more frequently. In contrast, sellable tokens are launched at a significantly lower rate. 

A key question is how many swaps newly created tokens experience on average and whether they sustain trading activity or quickly become inactive. This behavior differs between \textit{honeypots} and \textit{sellable} tokens. Since honeypots prevent selling, they often see initial high activity as traders unknowingly buy in before realizing that they cannot exit. In contrast, sellable tokens allow for continuous trading, potentially leading to longer lifetimes. Figure~\ref{fig:Num_Dist} confirms this: Honeypots exhibit a sharp peak at low swap counts before rapidly decaying, indicating short-lived activity. sellable tokens, while also peaking at low swaps, show a broader distribution, suggesting prolonged trading. This is evident in the median swap count, which is 42 for honeypots but significantly higher at 363 for sellable tokens. Figure~\ref{fig:AmountETHDist} presents the distribution in swap size in {W}ETH. 
The sell (negative) side appears uniform, while the buy (positive) side shows distinct peaks at specific {W}ETH values, likely due to bots executing trades with fixed investment amounts. 
This highlights that a substantial portion of market liquidity remains trapped in honeypots, misleading traders, and distorting the perceived trading volume. 
{It is worth noting that the sell-side transactions observed in honeypots typically originates from privileged wallets under the control of the token creator, who may execute or simulate sales to create the illusion of legitimate market activity.}

\section{A buy-and-hold strategy on new tokens}\label{sec:B&H}
In this section, we investigate the potential profitability of newly created tokens under the assumption that an investor is able to enter a position early in the lifecycle of the token, specifically, at the 60th swap. By systematically applying this strategy across the entire set of newly launched tokens, we can extract valuable insights into their financial behavior. 
{This choice of the 60th swap as entry point is motivated by structural and empirical considerations. 
Entering earlier, within the first block of transactions, would expose the investor to unrealistically high gas fees and blockchain congestion. 
Given that the average number of swaps per block at launch is around 27, the 60th swap allows us to avoid the firsts blocks, such that the participation becomes feasible and transaction costs are lower. 
Moreover, as shown in Figure~\ref{fig:Num_Dist}, most tokens—especially honeypots—cease trading activity after only a few swaps (median $\sim$42), so starting slightly later allows us to exclude the large fraction of tokens that “die” almost immediately. 
Finally, as detailed in Appendix~\ref{App:Returns}, we confirm empirically that the expected profit function is monotonically increasing when the entry threshold is set at the 60th swap, supporting this value as a reasonable choice. 
Of course, we acknowledge that the selection of 60 swaps retains a degree of arbitrariness. 
However, since profits are dominated by extreme honeypot returns, as we shown in this Section, varying the entry threshold within a reasonable range (e.g., 50–100 swaps) does not materially alter the qualitative results. 
Therefore, the 60-swap threshold can be regarded as a stable and representative benchmark for our analysis.}

As will become clear in the following, this analysis will help us reveal the profound impact of external factors that play a crucial role in decentralized exchanges, especially in the early stages of the lifecycle of a token. These factors, including the prevalence of honeypots, sandwich attacks, and liquidity manipulation, significantly alter the expected returns of our strategy, underscoring the need for a deeper understanding of the decentralized market structure for investment. 

The data reveal that tokens with less than 500 swaps constitute 90.6\% of the total, while those with less than 60 swaps make up 58.2\%. These numbers highlight a significant proportion of illiquid tokens. Given this observation, to make an exploratory study of the profitability, we  apply a simple buy-and-hold strategy. By using the constant product rule, we simulate the process of buying and selling tokens. Uniswap V2 operates on a constant product market maker model, where the relationship between reserves is given by
\ba{cpr}
    x \cdot y = k = L^2 \,\,,
\ea
where \(x\) and \(y\) are the reserves of two tokens in a liquidity pool, and \(k\) is a constant which defines the liquidity $L$ square. When tokens are swapped, this invariant slightly changes due to a 0.3\% fee, which is reinjected into the liquidity pool.

Our trading bot operates in swap time. In particular, it observes transactions up to the 60th swap before performing its first purchase of tokens using an initial investment of $\Delta x = 0.016$ {W}ETH, buying an amount $\Delta y$ of the new token computed using the constant product rule formula
\[
\Delta y = \frac{r \cdot y_{60} \cdot \Delta x}{x_{60} + r \cdot \Delta x},
\]
where \(x_{60}\) is the reserve of the newly created tokens in the liquidity pool at the 60th transaction, \(y_{60}\) is the correspondent reserve of {W}ETH in the liquidity pool and \(r = 0.997\) is the effective fee factor, accounting for the 0.3\% swap fee. 
{The initial investment \(\Delta x = 0.016\ \mathrm{ETH}\) was chosen based on the empirical distribution of swap volumes at token launch. 
This value lies close to the mode of the distribution and therefore represents a typical trade size observed on Uniswap~V2, ensuring a realistic order magnitude while keeping market impact reasonably small.}

After buying $\Delta x$ tokens, the strategy evaluates subsequent swap steps $n$ to determine an optimal time to sell. For each transaction $n>60$, the potential {W}ETH receivable from selling the tokens (\(\Delta x_{n}\)) is calculated as
\[
\Delta x_{n} = \frac{r \cdot x_n \cdot \Delta y}{y_n + r \cdot \Delta y},
\]
where \(x_n\) and \(y_n\) are respectively the token reserves of the token and {W}ETH in the pool at transaction $n>60$.
The revenue \(S_n\) at each transaction from the sale is computed as
\[
S_n = \Delta x_n\cdot r
\]
and the correspondent net profit (\(P\)) is
\[
P_n = S_n - \Delta x - 2G,
\]
where \(G\) is the gas fee incurred for each transaction that we take fixed to $G=0.002$ {W}ETH. 
{The value \(G = 0.002\ \mathrm{{W}ETH}\) corresponds to the average cost of a standard Uniswap~V2 swap consuming roughly \(1.1\text{–}1.5\times10^5\) gas at a price-per-gas of 12–20 gwei, 
which reflects typical post-launch network conditions.\cite{etherscanGasTracker,uniswapRouter02} }
This level implies an average fee of about 4–6~USD per swap (for {W}ETH $\simeq 2,500$~USD), consistent with historical gas costs on Ethereum. 
Since gas represents a fixed per-transaction cost, its relative impact on profitability rapidly diminishes for sufficiently large trades or returns, becoming negligible in percentage terms.
 The strategy uses predefined thresholds to decide the swap time $n$ when to sell the tokens:
\begin{enumerate}
    \item  For  \(n < 200\), the token is sold if \(S_n> 2  (I + 2G)\) ,
    \item For \(200 \leq n < 300\), the token is sold if \(S_n>1.5  (I + 2G)\),
    \item For \(300 \leq n < 400\), the token is sold if \(S_n>1.2  (I + 2G)\).
\end{enumerate}

If none of the above conditions are met and \(n \geq 800\), the bot sells if \(S_n > I + 2G\). If no sale occurs by \(n \geq 1500\), the strategy records a loss
\[
P_n=P_{loss} = -I - G.
\]
The same is accounted if the order flow of the token ends before one of these conditions is reached. {The choice of these thresholds is informed by the empirical trading lifetimes observed in Figure~1(a). 
While honeypots typically cease activity after only a few dozen swaps (median $\sim$42), sellable tokens display a median activity of about 363 swaps, and the vast majority stop trading by roughly the 400th swap. 
It is therefore reasonable to define selling windows within this range, progressively lowering the profit multipliers (2, 1.5, 1.2) as the swap count increases to emulate a decreasing risk appetite over time. 
These cutoffs are to some extent arbitrary, yet their impact on the aggregate outcomes is limited: 
given the extremely high returns and the dominance of honeypots that cannot be sold, moderate shifts in the thresholds (e.g., $\pm50$ swaps or $\pm0.2$ in the multipliers) do not materially affect the qualitative results.}
\begin{figure}[!htb]
  \centering
  \includegraphics[width=0.46\textwidth]{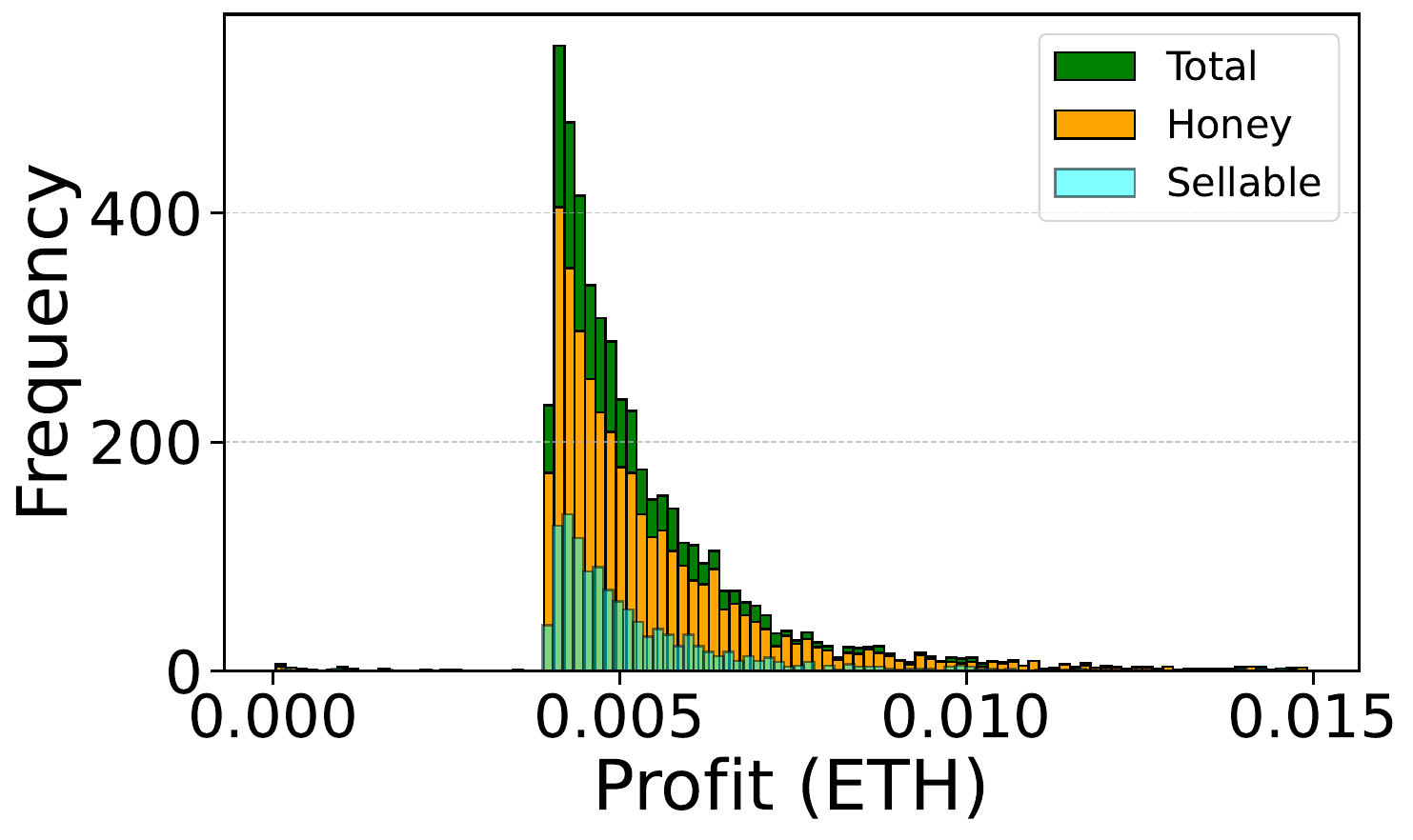}
  \caption{Distribution of profits in {W}ETH generated by our strategy. Among the 7,213 tokens in which the algorithm invested, 2,101 (\(\sim 29.1\%\)) failed to exit their position, resulting in a fixed loss of \( P_{\text{loss}} = -0.018 \) {W}ETH. All remaining tokens yielded a positive profit. To improve visualization, we restrict the distribution to the range between 0 and 0.015 {W}ETH. Beyond this threshold, there are an additional 162 tokens reaching a maximum profit of 1,056.07 {W}ETH.}
  \label{fig:profit_tot}
\end{figure}
We apply this strategy to all the coins. The overall combined results indicate that $\sim$ 70.9\% of the coins give a positive profit, while approximately \( 29.1\% \) of the tokens result in the fixed negative profit of \( P_{\text{loss}} = -0.018 \) {W}ETH. The profit distribution for all tokens is shown in Figure~\ref{fig:profit_tot}, with the range restricted to \( P \in [0,0.015] \) to improve visualization. We do not display the negative histogram at $P=P_{\text{loss}}$ and positive outliers. The maximum profit recorded on one coin is 1056.07 {W}ETH. The total profit over all the coins, namely
\[
\mathcal{P} = \sum_i P_i,
\]
reaches 2180.24 {W}ETH, with a total investment of 140.06 {W}ETH, leading to an impressive percentage profit of 1556.67\%. As expected, however, a large percentage of it is eaten by honeypots. 

Indeed, focusing specifically on the honeypot results, we observe an even higher positive percentage of 71.83\%, which includes the maximum profit of 1056.07 {W}ETH. The total profit reaches 2163.81 {W}ETH on a total investment of 104.88 {W}ETH, leading to a percentage profit of 2063.13\%. This clearly demonstrates that the majority of the apparent profit is generated by honeypot coins, creating the illusion of high profitability. However, this is entirely deceptive, as, despite holding assets that appear to have a significant market value at a given time, their actual worth is effectively zero since they cannot be sold. Thus, while these coins may seem extremely promising for a uninformed investor, they do not reflect a realizable profit in practice. Surprisingly, even with the sellable coins, the strategy shows positive results. Although the negative profit percentage is higher (31.98\%), the total profit still reaches 16.43 {W}ETH with a total investment of 35.18 {W}ETH, leading to a percentage profit of 46.69\%. Although significantly lower than honeypots, these results show that even a more conservative approach seems to yield profits. 
{We also verified the robustness of our results by varying the initial parameter space — including the entry swap threshold, gas fee assumptions, and exit conditions of the strategy. Despite these changes, the qualitative behavior of the results remains unchanged, owing to the exceptionally large returns observed across the tokens. We
also tested an extreme scenario with a gas fee of G = 0.008 ETH, corresponding to a network
price of approximately 66 gwei for a 120,000-gas swap—values typical of highly congested blocks
right after token deployment. Note that during the period considered in this paper the maximum average gas price in one day has been 52.65 gwei~\cite{etherscanGasTracker,uniswapRouter02} . Even under these extreme conditions,
the overall percentage profit remains very large, reaching about 1074\%. We also note that small variations of this threshold (e.g., 50 or 100 swaps) do not materially affect the qualitative results, as the average profits shown in Figure~2 remain extremely large. In particular, using the same inputs, the average return is approximately 722\% when entering at the 50th swap, 575\% at the 70th, and 495\% at the 100th. Hence, although the 60th swap provides a theoretically and empirically sound baseline, varying this value leads to very similar conclusions.}
At this point, a spontaneous question arises: are we missing something, or is it really this easy to profit from the creation of new coins, even with a simple strategy, provided that we are (which is not trivial) able to distinguish between honeypots and sellable coins?

\section{The limits of new coins profitability}\label{sec:Limits}
In the previous section, we assessed whether a simple buy-and-hold strategy could generate profit in newly created tokens. To further investigate, we analyze key limitations that significantly reduce its practical profitability.

Three main factors constrain this strategy. The first is the prevalence of honeypots, which prevent selling and drastically limit tradable opportunities. To quantify their impact, we introduce \textit{Net Traded Value} (NTV), measuring the value trapped in honeypots and its effect on market liquidity. NTV also highlights two additional risks: large liquidity withdrawals by token owners causing price crashes and rug pulls, both of which contribute to investment losses.

Even if one could reliably distinguish honeypots, another major limitation arises from sandwich attacks. Our strategy sells after the front-runner’s buy in a sandwich attack, which appears effective in \textit{swap time} but is not possible in \textit{physical time}. Successfully executing this requires precise transaction positioning within the same block, which is difficult due to blockchain mechanics and competition in the mempool. Additionally, most sandwich attacks occur via private relays like \href{https://flashbots.net/}{FlashBots}, making them invisible in the public mempool and harder to exploit without specialized access.

This leads to the third limitation: the difference between price evolution in swap time and physical time. Because transactions are executed in discrete blocks, the apparent profitability observed in swap time may not hold in real-world trading. To study this, we apply clustering techniques to analyze price trajectories in both temporal frameworks.

The following sections provide a detailed quantitative assessment of these limitations and their implications for trading in newly created tokens.

\subsection{Net Traded Value as a measure of market value}
To quantify the financial impact of newly created tokens \cite{experimental}, we introduce the concept of NTV. 
This metric allows us to quantify the value {of} these tokens, providing a clearer picture of the actual economic significance of new token creation. 
In particular, by distinguishing between honeypots and sellable tokens, NTV helps reveal that a substantial portion of market liquidity is, in reality, fictitious, as it remains trapped in honeypots. 
This distortion makes the market appear more valuable than it actually is, highlighting the discrepancy between perceived and real liquidity. 
Moreover, token NTV allows us to identify additional risks associated with newly created tokens, particularly two major pitfalls: the liquidation of a large portion of liquidity by the owner, leading to drastic price collapses, and rug pulls, where liquidity is removed from the pool. 
In the latter case, investors are left holding tokens that, despite appearing to have a certain market value, are effectively worthless due to the complete absence of liquidity, making them impossible to sell. 

{From an economic perspective, NTV is not intended as an absolute performance metric but as a proxy for the cumulative economic value extracted from a liquidity pool over time, expressed in WETH. 
By aggregating directional activity (inflows minus outflows) in value terms, NTV captures the net flow of economic value rather than raw traded volume, making it possible to distinguish between transient swap activity and persistent capital outflows linked to arbitrage or manipulation. 
Although our dataset does not include all pools or Uniswap versions, NTV remains internally consistent at the pool level, since MEV and arbitrage cycles are typically local and self-contained. Furthermore, as discussed earlier, the presence of parallel pools for tokens originally created on Uniswap~V2 across other DEXes is negligible, and thus does not materially affect the interpretation of NTV.
Consequently, relative variations in NTV provide an indication of trading pressure and value extraction dynamics within each observed market.}

We define the Net Traded Value (NTV) as
\begin{equation}
{\rm \text{NTV}}^i(N_t) = \left( \sum_{t=1}^{N_t} v^i_t \right) \cdot p^i(t=N_t)
\end{equation}
where \( v^i_t \) represents the signed token amount exchanged in the \( t \)-th swap for the \( i \)-th token, \( N_t \) is the number of swaps considered, and \( p^i(t=N_t) \) denotes the token marginal price (in {W}ETH) of the new coin at the time of the \( N_t \)-th swap for the \( i \)-th token.

\begin{figure}[!htb]
\centering
\begin{subfigure}{0.46\textwidth}
  \centering
  \includegraphics[width=\linewidth]{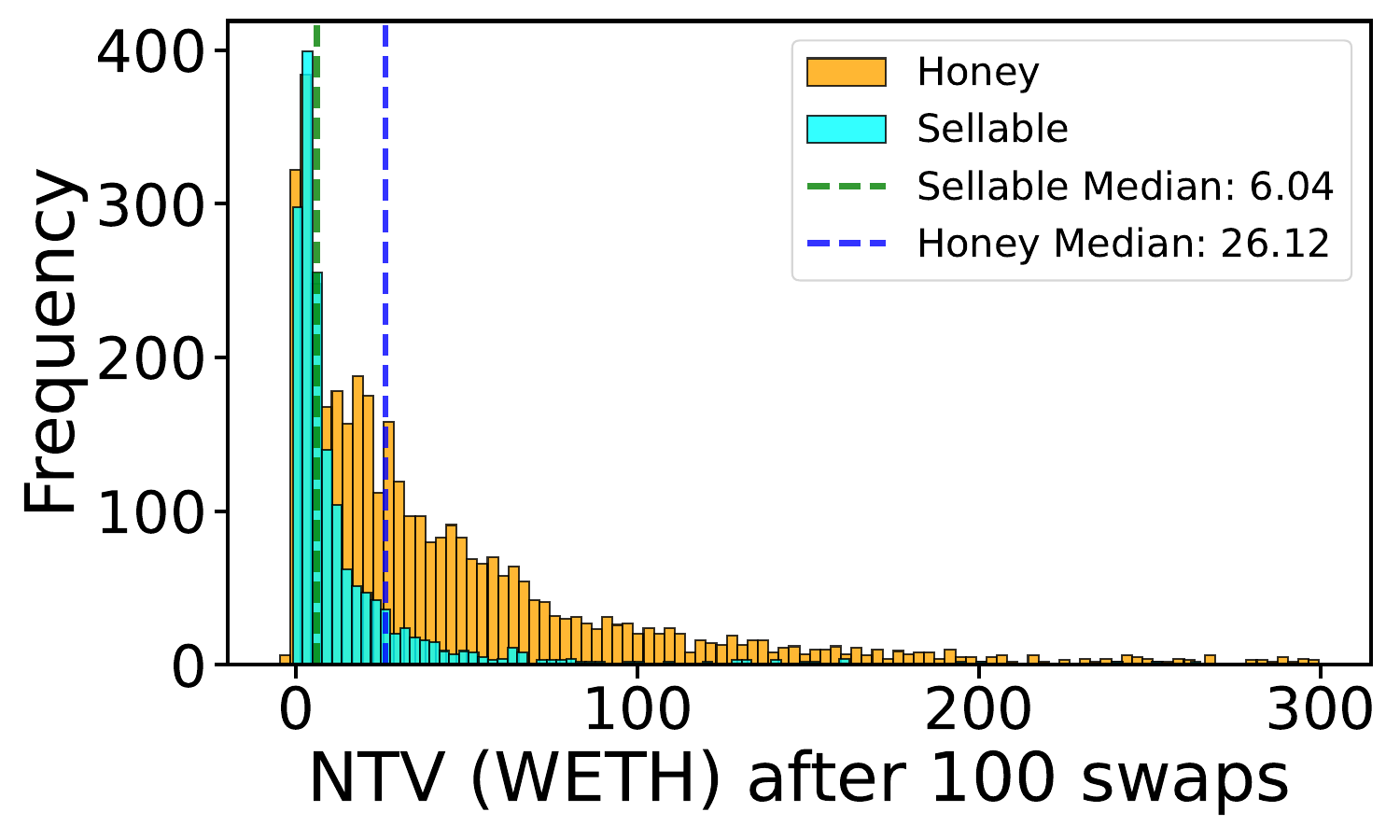}
  \caption{Histogram of NTV after 100 swaps. The minimum value for sellable tokens is -22.06 {W}ETH while the maximum is 1080.75 {W}ETH. Instead for honeypots are respectively $-3.922 \times 10^{13}$ {W}ETH and $2.096 \times 10^{13}$ {W}ETH.}
  \label{fig:Histo_Cap_100}
\end{subfigure}
\hfill
\begin{subfigure}{0.46\textwidth}
  \centering
  \includegraphics[width=\linewidth]{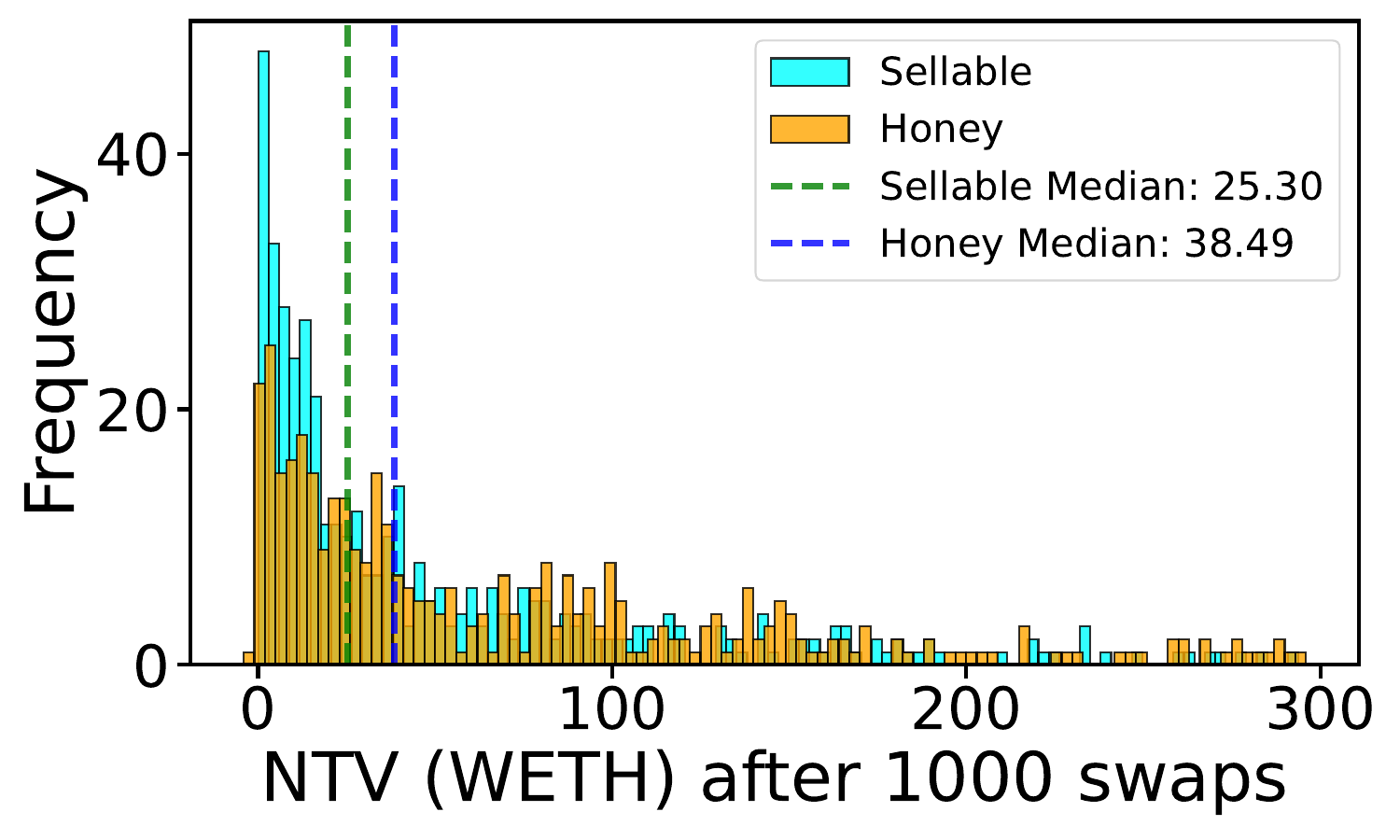}
  \caption{Histogram of NTV after 1000 swaps. The minimum value for sellable tokens is -16.30 {W}ETH while the maximum is 5787.21 {W}ETH. Instead for honeypots are respectively -14.55 {W}ETH and $1.331 \times 10^5$ {W}ETH.} 
  \label{fig:Histo_Cap_1000}
\end{subfigure}

\caption{Histograms of the NTV of the new tokens after different number of swaps.}
\label{fig:Istogrammi_Cap}
\end{figure}

In Figure~\ref{fig:Istogrammi_Cap} we show the {distribution of} NTV of new tokens for $N_t = \{100,1000\}$ for both sellable and honeypot tokens. 
\begin{figure}[!htb]
    \centering
    \includegraphics[width=10cm]{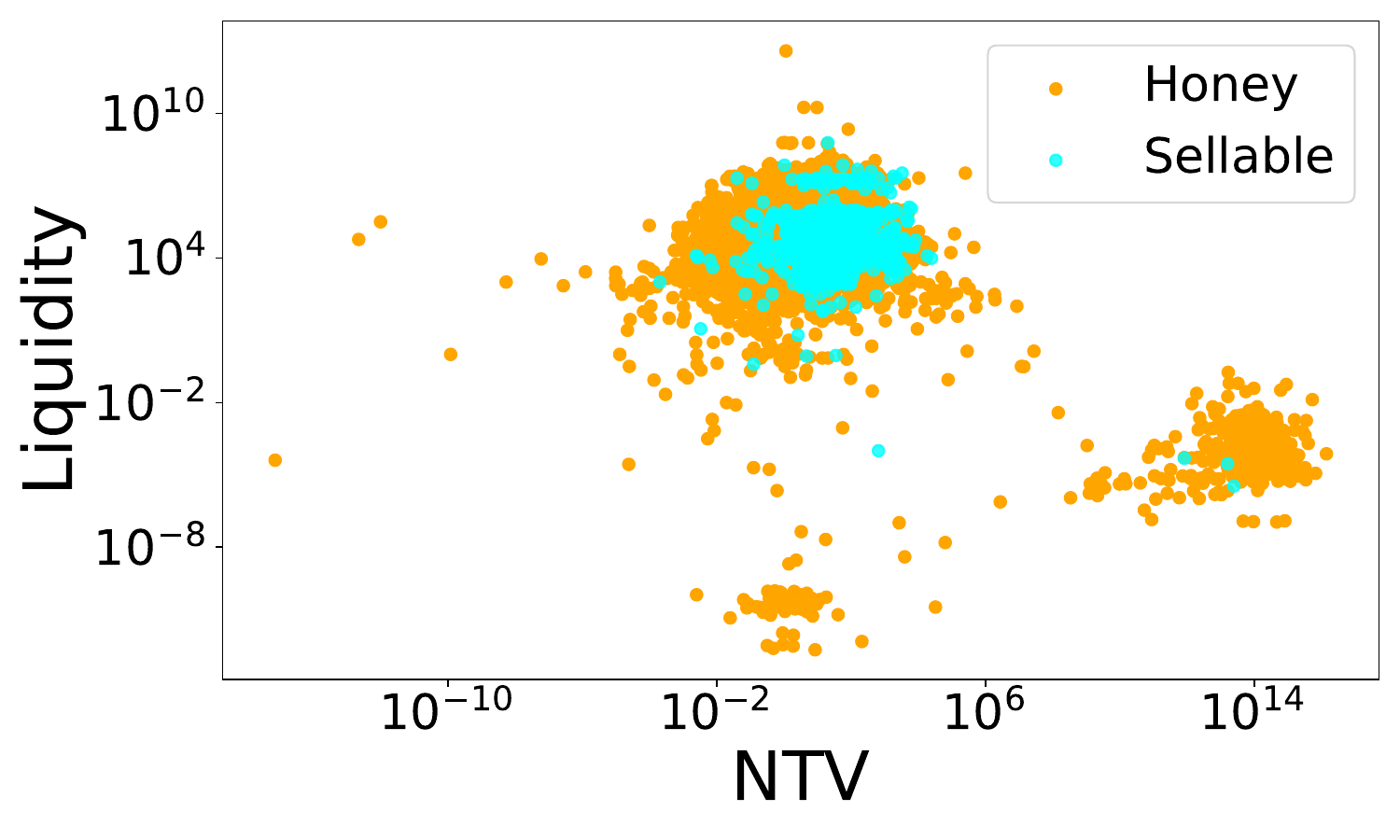}
    \caption{Scatter plot showing the maximum NTV value in the lifetime of all new tokens against the corresponding {l}iquidity value at that time. The $0.14\%$ of tokens with a maximum negative NTV are excluded. A negative NTV occurs when a hidden supply held by the token creator, which was not included in the initial mint, is used to sell into the pool. }
    \label{fig:scatter_plot_liq_TK}
\end{figure}
The first noticeable observation is that the number of tokens present in the histograms decreases rapidly, especially for honeypots, as expected from the trends already observed in Figure~\ref{fig:Num_Dist}. As the number of swaps increases, the distribution of honeypots becomes increasingly comparable to that of sellable tokens. We define the total new coins market value at fixed swap number $N_t$ as the sum of all the  Net Traded Values (NTVs) of the new coins at that time, namely
\begin{equation}
 \text{NTV}_{\text{tot}}(N_t) = \sum_i {\rm \text{NTV}}^i(N_t). 
\end{equation}
Table~\ref{tab:capitalization} reports the results distinguishing between honeypots and sellable tokens. We observe that at the 10th swap, 94\% of the new token market value is trapped in honeypots, meaning that the actual total NTV of sellable tokens after 10 swaps accounts for only 6\% of the observed total. As the number of swaps increases, the NTV of honeypots gradually becomes comparable to that of sellable tokens. However, even at the 1000th swap, more than half of the total new NTV remains locked in honeypots, highlighting the persistent impact of these deceptive schemes on market liquidity.
\begin{table}[h]
    \centering
    \begin{tabular}{c|c|c}
        \hline
        \textbf{$N_t$} & \textbf{honeypot $\text{NTV}_{\text{tot}}$ ({W}ETH)} & \textbf{sellable $\text{NTV}_{\text{tot}}$ ({W}ETH)} \\
        \hline
        10 & 16857.19 (94\%) & 1059.91 (6\%) \\
        100 & 170150.11 (86\%) & 27450.66 (14\%) \\
        1000 & 25485.46 (55\%) & 20867.45 (45\%) \\
        \hline
    \end{tabular}
    \caption{Amount of NTV of honeypots and sellable tokens at different swap thresholds.}
    \label{tab:capitalization}
\end{table}

Interesting insights can be extracted from analyzing extreme values of NTV, both negative and positive. Let us start with the \textit{negative ones}. Indeed, as previously discussed, NTV incorporates the cumulative \textit{signed} volume of recorded trades in its definition. When a token is created, the owner deposits an initial amount of liquidity into the pool, often using all or a portion of the total token supply. However, it is possible for the owner to retain a significant share of the token supply in their personal wallet. At any point, the owner may decide to sell these retained tokens in large quantities, causing the cumulative trade volume to turn negative and, in turn, to {obtain} a negative NTV. After this action the price collapses, leaving investors with a token whose value has dramatically declined. The negative values primarily originate from honeypot tokens.

\begin{table}[h]
    
    \footnotesize
    \setlength{\tabcolsep}{5pt}
    \renewcommand{\arraystretch}{1.2}
    \centering
    \begin{tabular}{c|c|c|c|c|c|c}
        \toprule
        \textbf{Action} & \textbf{Timestamp} & \textbf{Amount {W}ETH} & \textbf{Amount NT} & \textbf{Reserve {W}ETH} & \textbf{Reserve NT} & \textbf{Price ({W}ETH)} \\
        \hline
        Sell & 2024-11-01T20:52:47 & -0.02365 & 4.86E+09 & 0.89905 & 1.89E+11 & 4.75E-12 \\
        Burn & 2024-11-01T20:54:11 & -0.89905 & -1.89E+11 & - & - & - \\
        Buy & 2024-11-01T21:10:35 & 0.04965 & -1.46E-05 & 0.04965 & 0.0001165 & 426.19 \\
        \bottomrule
    \end{tabular}
    \caption{Sequence of transactions involving swaps and a burn event for the pair address 0xeeb1aec10fa7bae35716166199a812baf29fa2fc. It can be observed that the owner burns nearly all the liquidity (rug pool). Although not directly visible in the table, some residual liquidity remains after the burn, allowing the last swap to take place. Due to the drastically reduced liquidity, the price experiences a sharp increase, highlighting the significant impact of low liquidity on price dynamics.}
    \label{tab:swap_burn_sequence}
\end{table}

In contrast, extreme positive values are primarily associated with rug pulling events. Here, the owner, often the sole liquidity provider, removes all liquidity from the pool. This is possible if the owner does not impose restrictions on liquidity withdrawal. A sudden and significant liquidity reduction results in extreme price slippage, causing the token price to skyrocket even with a small buy order. An example is shown in Table~\ref{tab:swap_burn_sequence}. However, shortly after, the liquidity pool is extremely low, leaving investors holding tokens with an artificially inflated price but no means of selling them. In Figure~\ref{fig:scatter_plot_liq_TK}, we present a scatter plot on a log-log scale, showing the maximum NTV value in the lifetime of all new tokens against the corresponding liquidity $L$, as defined in Eq.~\eqref{cpr}, which corresponds to the point of extreme NTV in the case of a rug pull. A clear clustering pattern emerges, where the highest NTV values are concentrated around extremely low liquidity levels. This confirms that tokens reaching peak NTV often do so under conditions of minimal liquidity, reinforcing the idea that these extreme price surges are artificially induced by liquidity withdrawal strategies rather than genuine market demand. An important observation is that nearly all tokens clustered in the low-liquidity, high-NTV region are labeled as honeypots. This is likely due, in part, to the ability of honeypot schemes to attract numerous investors, from whom funds can later be extracted through rug pulls. However, this classification may also be influenced by the methodology used by our providers to distinguish between honeypots and sellable tokens. For instance, honeypot.is identifies honeypots by simulating sell transactions, which may fail even in the case of a rug pull, without necessarily requiring explicit honeypot code in the contract.

These extreme cases, both negative and positive, highlight the various forms of market manipulation present in newly created tokens, further emphasizing the risks associated with investing in such projects.

\subsection{Sandwich attacks}
Using our strategy, only 0.28\% of sellable tokens generate a profit exceeding 1 {W}ETH. Excluding these outliers, the cumulative profit turns negative, indicating that success relies on a few tokens with extreme gains. This raises the question: what drives this small fraction of highly profitable trades? The answer lies in \textit{sandwich attacks}.

A sandwich attack exploits price slippage by front-running a victim buy order. The attacker detects the victim trade in the mempool, submits a buy with a higher gas fee to execute first, inflating the price, and then sells immediately after the victim purchase. This process is optimized using \href{https://flashbots.net/}{FlashBots}~\cite{flashbot2022}, which bypasses the public mempool to prevent competition and ensure precise execution~\cite{MEV, mitigationMev} by submitting transaction bundles. Uniswap AMM model and low-liquidity pools amplify these attacks, as even small buy orders can cause significant price slippage. We estimate from our data that, among tokens with at least 100 swaps, approximately $68\%$ of honeypot tokens and $90\%$ of sellable tokens have experienced at least one sandwich attack event\footnote{We define a sandwich attack event as one in which, at the same timestamp (within the same block), two swaps of opposite signs occur, where the absolute value of the swapped token amounts are similar within a $5\%$ threshold.}.

Our strategy takes advantage of some of these events, which are theoretically profitable when considering price evolution at swap time but not in reality. Exploiting these events generate the small percentage of extreme positive profits that, at least theoretically, make our strategy profitable overall. As sandwich attackers typically execute large trades relative to liquidity, the resulting price spike is substantial. By systematically capturing these price distortions, our approach remains profitable. Sharp price peaks followed by immediate drops correspond to sandwich attacks, affecting both sellable tokens and honeypots. Our strategy exploits these peaks, securing profits before price reverts. Without these attacks, the strategy would yield negative returns and, thus, in real investments would not be exploitable to obtain positive profits.  

This further observation reveals that profitability with our strategy is not just limited by the presence of honeypots. Even if one could perfectly filter them out, executing profitable trades would still require precise positioning within an ongoing sandwich attack in the same block. This is nearly impossible, as most of these attacks use FlashBots, bypassing the public mempool. Our findings emphasize the need to consider \textit{physical time} rather than relying solely on \textit{swap time}. Transactions on the blockchain are processed in discrete blocks (typically 12 seconds apart), where multiple trades, including sandwich attacks, execute simultaneously. To exploit this strategy in practice, one would need to ensure their sell order is placed immediately after the front-runner buy—an extremely challenging task.

The distinction between swap time and block time not only complicates strategy execution but also impacts the study of new coin price behavior. This issue is further analyzed in the next section using clustering techniques.

{\bf Optimal Sandwich Strategy and its higher profitability in low-liquidity pools.} Determining the optimal investment amount for a sandwich attack is crucial\footnote{In Ref.\cite{heimbach2024sandwich}, the authors derive the optimal strategy in the limit of huge liquidity or when the pool fee is zero ($r = 1$).}. Indeed, while increasing the investment might seem beneficial, in the limit of an infinitely large trade, the attacker would effectively perform a round-trip trade that results in losses due to pool fees and gas costs, without generating any real profit. Therefore, the optimal investment size must balance maximizing the price impact from the front-running trade while avoiding excessive slippage and transaction costs. 

Introducing $s$ as the gain in WETH for the sandwich attacker, we define the maximum value $s_{\rm max}$ as occurring when the amount of the WETH swapped by the attacker in the first transaction, $\Delta x_a^{\rm max}$ , is (see Appendix \ref{appendix:sandwichAttack} for details):
\begin{equation}
\label{eq: MaxPoint}
\Delta x_a^{\rm max} = \frac{ \Delta x_\epsilon }{f} - x, \,\,
\end{equation}
{where \( \Delta x_\epsilon \) is the amount of Ethereum sold by the  the victim and $x$ is the Ethereum reserve into the pool.} Expressing $s_{\rm max}$ using the value of $\Delta x_a^{\rm max} \,$:
\begin{equation}\label{eq: Maximum}
s_{\rm max} = \frac{ ( \Delta x_\epsilon - fx )^2 }{ \Delta x_\epsilon } \,\,.
\end{equation}
Therefore, since the existence condition $\Delta x_a \ge 0$, we have a positive gain if and only if $ \Delta x_\epsilon > fx$. This shows that there is not always an opportunity for a profitable sandwich attack \textcolor{blue}.

A crucial insight from this analysis is that performing a sandwich attack in a highly liquid pool, i.e., with large \( x \), leads to diminishing or even negative returns. This becomes clearer by taking the limit \( x \rightarrow \infty \) in Equation~\eqref{eq: Completes}, yielding:
\begin{align}
    \centering
    s(\Delta x_a;\Delta x_\epsilon) \rightarrow \Delta x_a (r^2-1) < 0, \quad \text{for } x \rightarrow \infty.
\end{align}
This result stresses a fundamental conclusion: sandwich attacks are far more profitable in low-liquidity pools. The reason is that in high-liquidity pools, the price impact of a single trade is much lower, making it harder for the attacker to extract significant profit from the victim trade. This insight may explain why sandwich attacks are particularly prevalent in newly created tokens, which typically have very low liquidity, making them ideal targets for such exploitative strategies.

When executing a trade on Uniswap V2, users can set a \textit{slippage tolerance}, i.e. $\eta \in [0,1]$, which defines the maximum acceptable price deviation from the expected execution price. This protects traders from excessive price impact caused by sudden market fluctuations or front-running attacks. In the context of sandwich attacks, this means that the victim effectively imposes an upper limit on the price they are willing to pay for the token. This corresponds to placing a vertical line on the plot shown in Figure~\ref{fig:Max_Sandw}, intersecting the profit curve. If this line is positioned below the peak, the optimal strategy for the attacker is to invest precisely the amount that corresponds to the victim slippage tolerance. An interesting property of this scenario is the symmetry around the peak, implying that two different investment amounts—one smaller and one larger—can yield the same profit. On the other hand, if the slippage tolerance is higher than the peak, the attacker optimal profit is achieved at the peak itself. 


\subsection{The different behavior of new coins in swap and physical time}

In the previous sections, the crucial distinction between \textit{swap time} and \textit{physical time} when analyzing price dynamics on Uniswap emerged. This differentiation is particularly relevant when evaluating the feasibility of profit opportunities, as real-world execution is constrained by block times and transaction ordering within the blockchain. In this section, we further investigate this aspect by studying the price evolution of newly created tokens during the initial phase of their existence. To achieve this, we analyze price trajectories using both swap time and physical time, applying \textit{clustering techniques} \cite{ClusterData} to identify recurring patterns and behaviors in the early life cycle of newly created tokens and providing a comparative view of how their prices evolve under these different temporal frameworks.
By grouping tokens with similar price evolution, we can better understand the common trends of honeypots and sellable coins. Additionally, this clustering enables us to assess how different price behaviors influence potential profitability depending on whether trades are executed based on swap events or real-world block time constraints. As we will see, the findings reinforce the conclusions from the previous paragraphs regarding the limitations of our trading strategy. The discrepancies between swap time and physical time play a fundamental role in determining the actual accessibility of profit opportunities, as well as the risks associated with different trading approaches.
{A detailed discussion of the clustering results, including the elbow curve and PCA visualizations for $k=2$ and $k=3$ clusters, is provided in Appendix~\ref{app:cluster}.}

\begin{figure}[!htb]
\centering
\begin{subfigure}{0.46\textwidth}
  \centering
  \includegraphics[width=\linewidth]{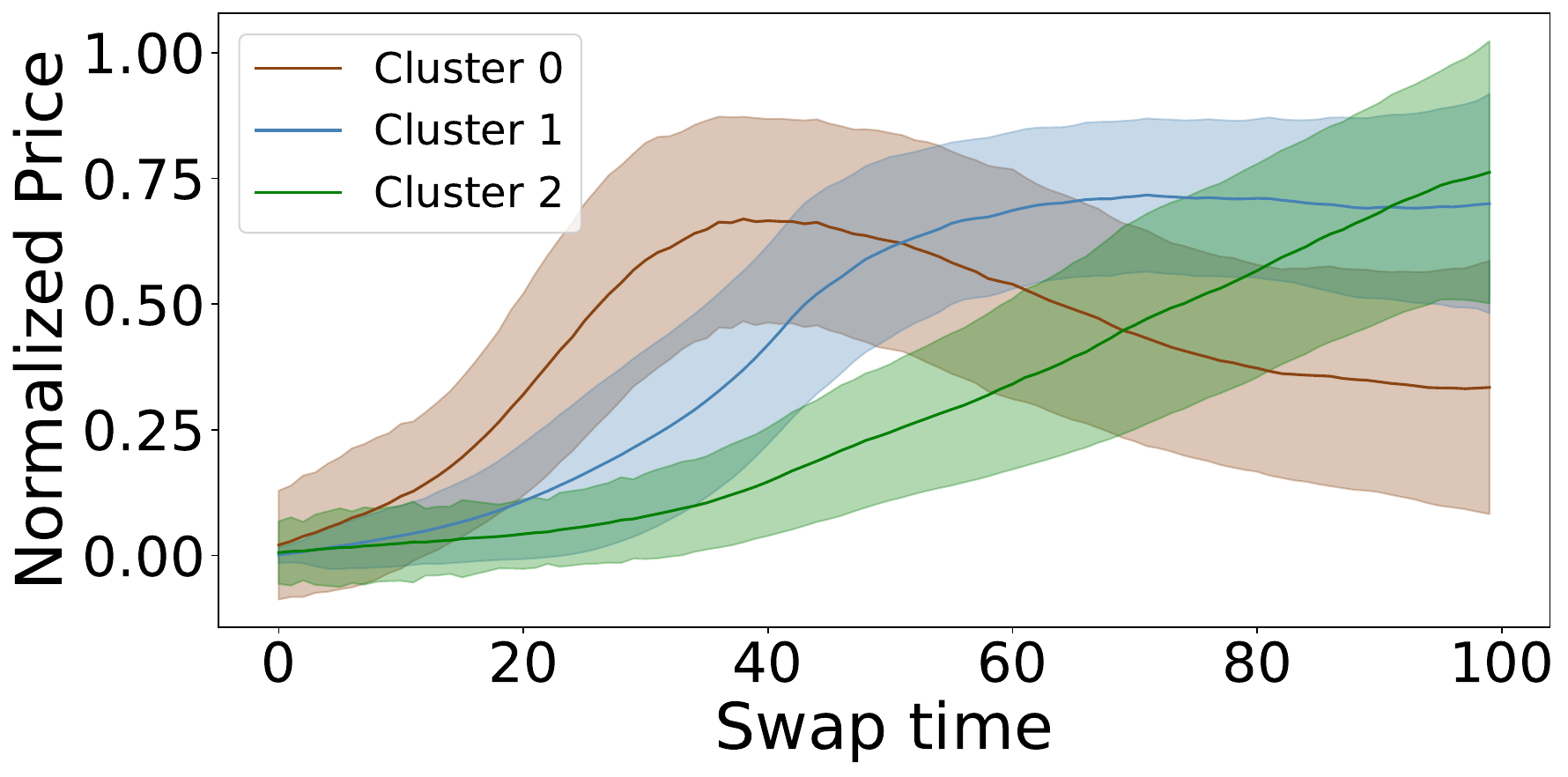}
  \caption{Clustering resulting trajectories for $k=3$.}
  \label{fig:Tr_HS_Sw}

\end{subfigure}
\hfill
\begin{subfigure}{0.46\textwidth}
  \centering
  \includegraphics[width=\linewidth]{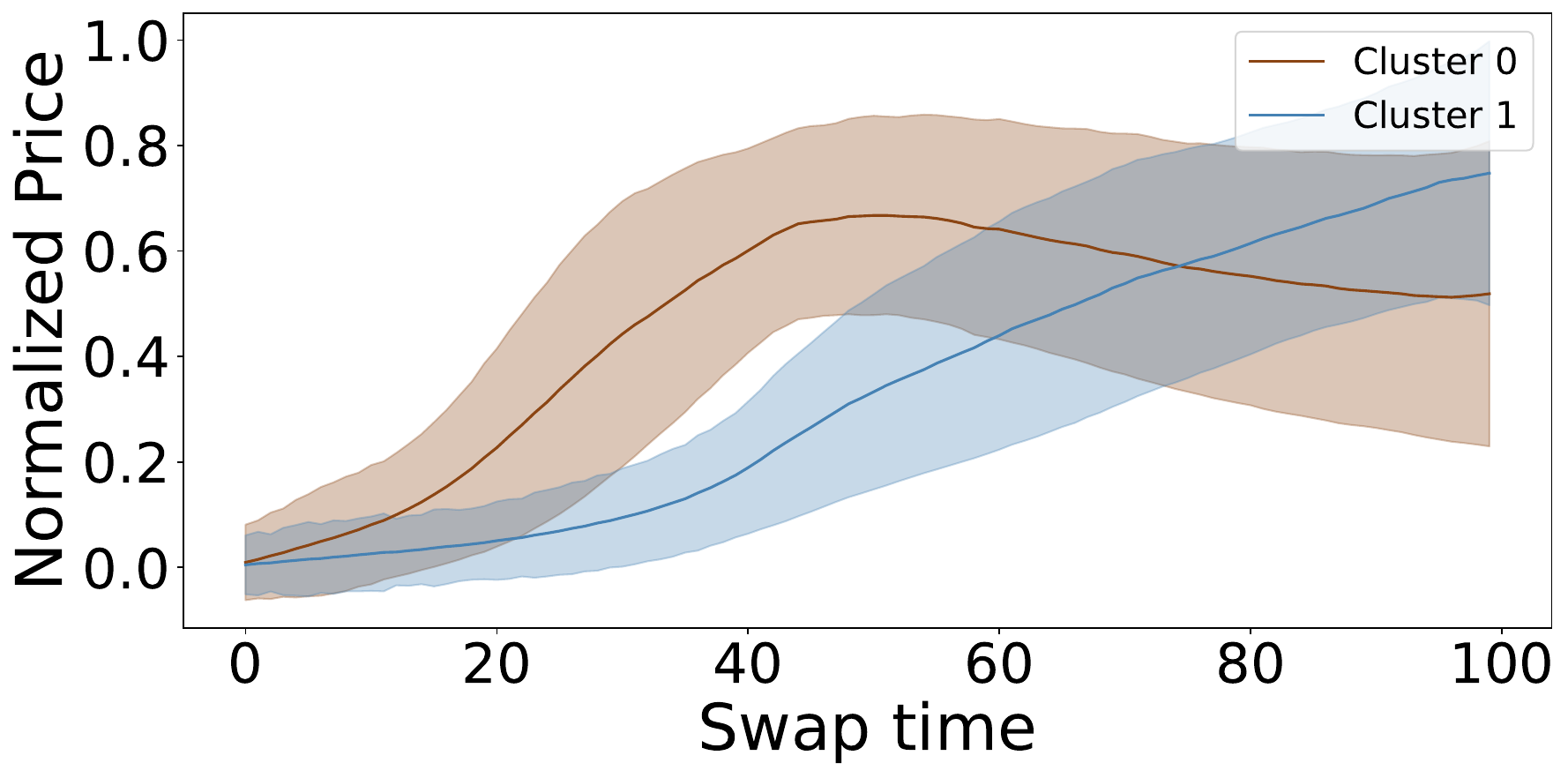}
  \caption{Clustering resulting trajectories for $k=2$}
  \label{fig:Tr_HS_2_Sw}

\end{subfigure}
\caption{The figures show the mean and the one standard deviation bands of the trajectories resulting from the clustering, with $k=2$ and $k=3$ clusters in swap time.}
\label{fig:Tr_HS_Tot}
\end{figure}

{\bf Clustering in swap time.} Let us first discuss the case in which time is defined in swap transactions. In this framework, we consider time series containing the first $10^2$ and $10^3$ swaps. Tokens that do not reach the selected number of swaps are discarded. To allow meaningful comparisons across different tokens, we normalize each price trajectory using a min-max scaler, which rescales values between 0 and 1. Once the price time series are normalized, we perform clustering using the K-Means algorithm. We set the number of clusters equal to two and three. In Figure~\ref{fig:Tr_HS_Sw} we show the resulting trajectories. The line represents the mean value, while the bands include one standard deviation.  The clustering results indicate the following distribution of trajectories among the three categories:  2459 coins in cluster 0;  2187 coins in cluster 1; 1195 coins in cluster 2. Thus, the distribution is pretty homogeneous among the three categories.  

Given our interest in distinguishing honeypot tokens from sellable ones in the early stages of their lifecycle, we set $k = 2$ to evaluate whether clustering can effectively separate these two categories by grouping a higher proportion of one type within each cluster. In Figure~\ref{fig:Tr_HS_2_Sw} we show the resulting trajectories corresponding to the two clusters.  In this case, the cluster distribution is more concentrated in cluster 1, with 3309, than in cluster 0, with 2532. In Table~\ref{tab:cluster_distribution} we show the distribution of the honeypots and sellable coins. We clearly observe a dominant presence of honeypots in cluster 1. This is expected, as honeypot tokens prevent selling, leading to a characteristic initial price behavior. In particular, since traders can only buy but not sell, the price tends to rise continuously due to persistent buy pressure without any corrective sell-side activity. This artificial price inflation creates the illusion of profitability, which is entirely misleading since these tokens cannot be liquidated.  In contrast, sellable tokens are significantly less frequent in cluster 1, suggesting that the price patterns captured in this cluster—likely a continuous upward trend—are primarily driven by the inability to sell in honeypots. In cluster 0, where both sellable and honeypot tokens are present in more comparable numbers, the price dynamics might be more balanced, reflecting a more natural market behavior where both buying and selling contribute to price formation. In this case, indeed, after an initial increase prices stabilize to a more reasonable behavior.
\begin{table}[h]
    \centering
    \begin{tabular}{c|cc}
        \hline
        \textbf{Cluster} & \textbf{honeypots} & \textbf{sellable} \\
        \hline
        0 & 24\% & 19\% \\
        1 & 46\% & 11\% \\
        \hline
    \end{tabular}
    \caption{Distribution of honeypots and sellable tokens across clusters for 100 swaps.}
    \label{tab:cluster_distribution}
\end{table}

Increasing the time-series length to $10^3$ swaps, we reduce the number of price trajectories we analyze because we exclude the coins whose activity ended before a thousand swaps. The trajectories resulting from the clustering are shown in Figure~\ref{fig:Tr_HS_Tot}. Taking $k=3$, there are  354 trajectories in cluster 0,  361 in cluster 1, and 242 in cluster 2. Decreasing to $k=2$, we find 534 trajectories in cluster 0 and  423 in cluster 1. 

\begin{figure}[!htb]
\centering
\begin{subfigure}{0.46\textwidth}
  \centering
  \includegraphics[width=\linewidth]{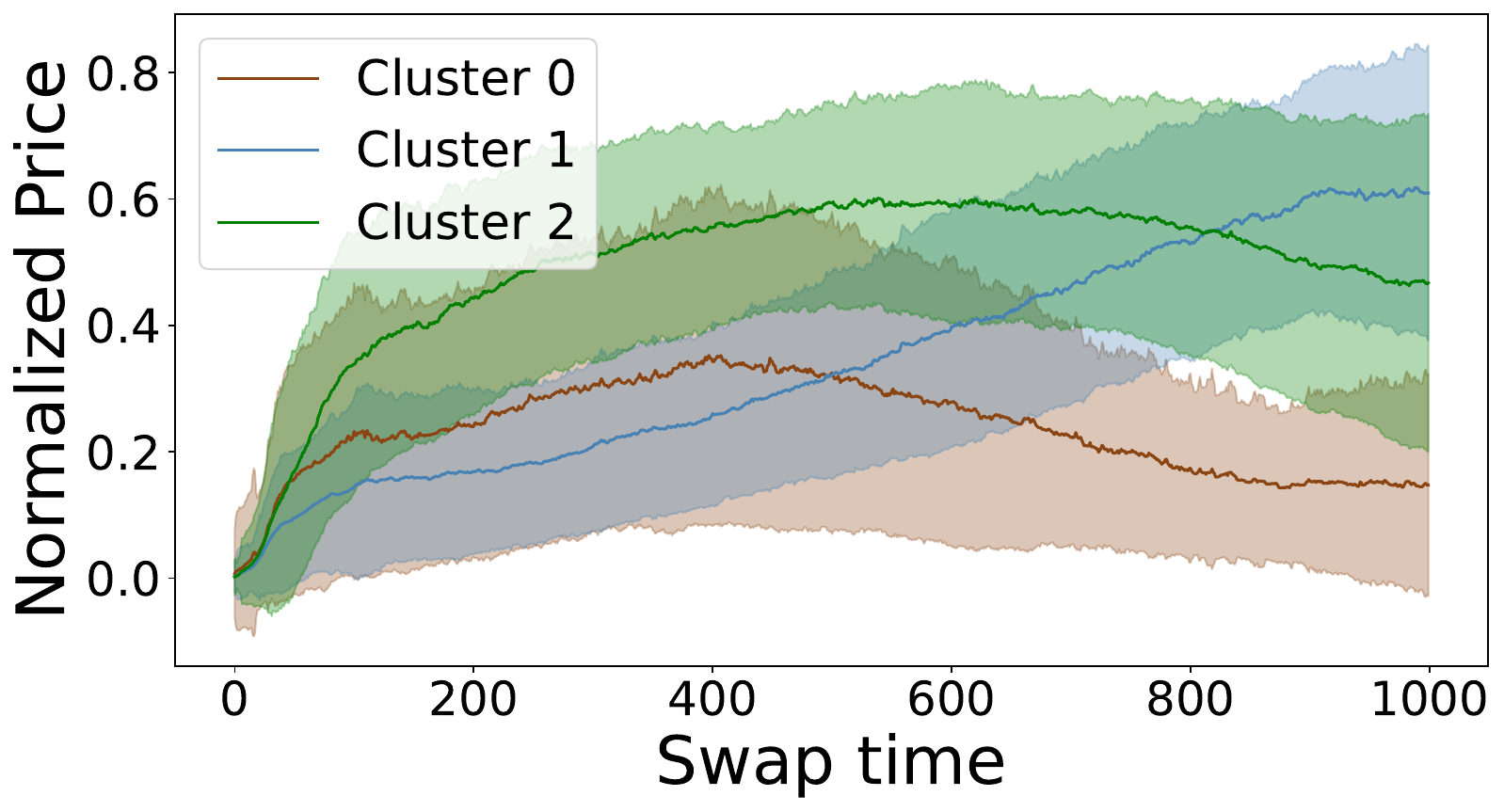}
  \caption{Clustering resulting trajectories for $k=3$.}
  \label{fig:Tr_HS}

\end{subfigure}
\hfill
\begin{subfigure}{0.46\textwidth}
  \centering
  \includegraphics[width=\linewidth]{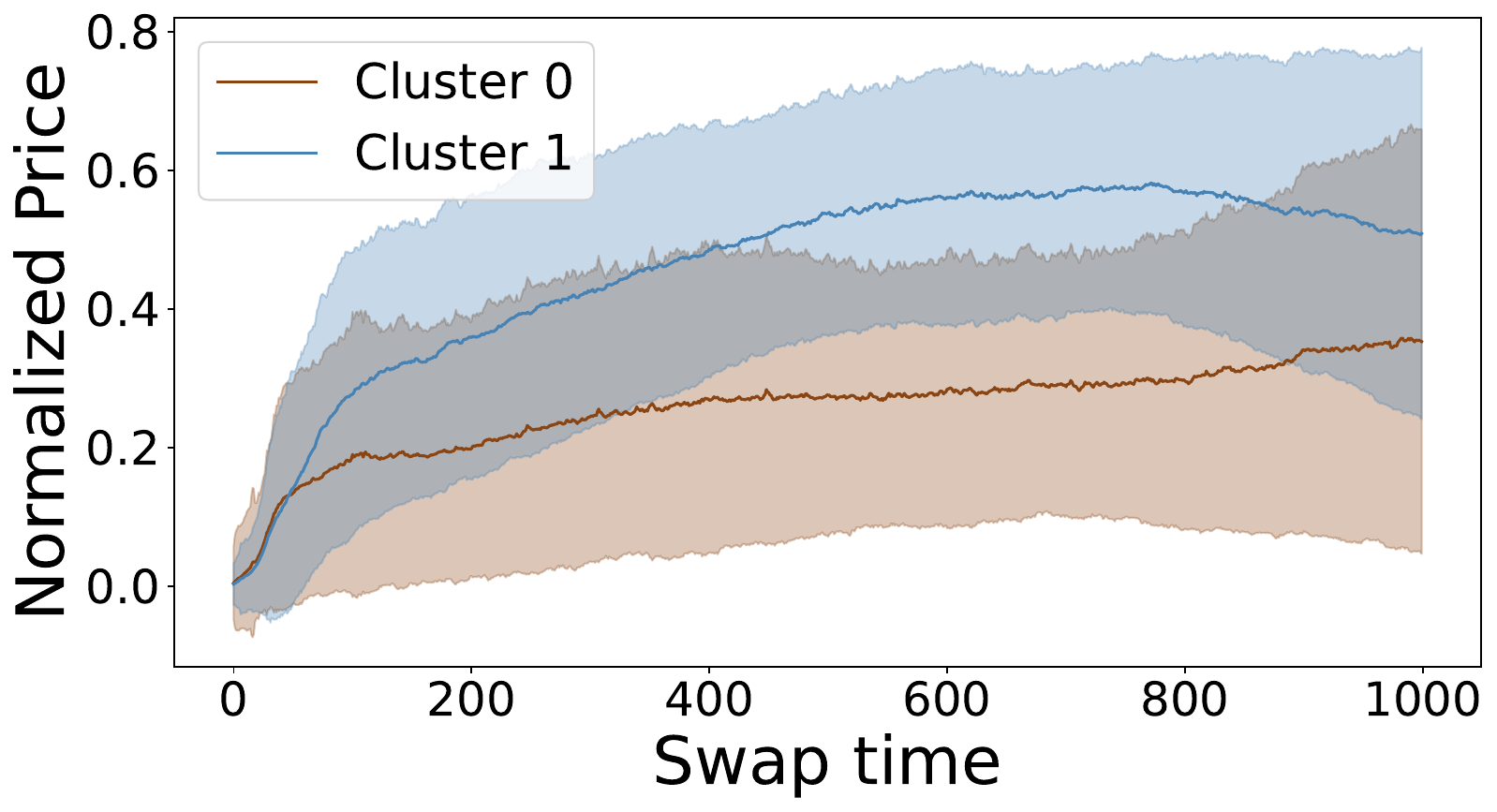}
  \caption{Clustering resulting trajectories for $k=2$}
  \label{fig:Tr_HS_2}

\end{subfigure}
\caption{The figures show the mean and the one standard deviation bands of the trajectories of length $10^3$ resulting from the clustering, with $k=2$ and $k=3$ in swap time.}
\label{fig:Tr_HS_Tot}
\end{figure}
Adopting clustering appears to reveal distinct and well-defined patterns, particularly for \( k=3 \), which may conceal important properties of the tokens. However, the distribution of honeypots and sellable tokens across the clusters, both for \( k=2 \) and \( k=3 \), is less clear, as they appear uniformly spread. This is likely due to the fact that, as the number of swaps increases, the proportion of honeypots compresses toward that of sellable tokens. In this sense, it is reasonable that they exhibit similar patterns over larger timescales.

In swap time, it appears from the clustering results that, on average, price trajectories tend always to increase in the first phase of a token’s lifecycle while a part of them, e.g. cluster 0 in Figure~\ref{fig:Tr_HS}, at a certain time reverse to lower values. However, it is important to consider that this observation is inherently biased by the fact that all clustering analyses in swap time are conditioned on tokens having survived at least the number of swaps considered. Moreover, this analysis does not take into account the fact that many of these transactions actually occur simultaneously due to the block-based structure of the blockchain. These limitations highlight the need for a more comprehensive approach. A deeper understanding of these dynamics can be achieved by performing the clustering analysis in physical time rather than in swap time, allowing us to capture the complete temporal structure of price evolution of all the tokens.

\begin{figure}[!htb]
        \begin{subfigure}{0.46\linewidth}
            \includegraphics[width=8cm]{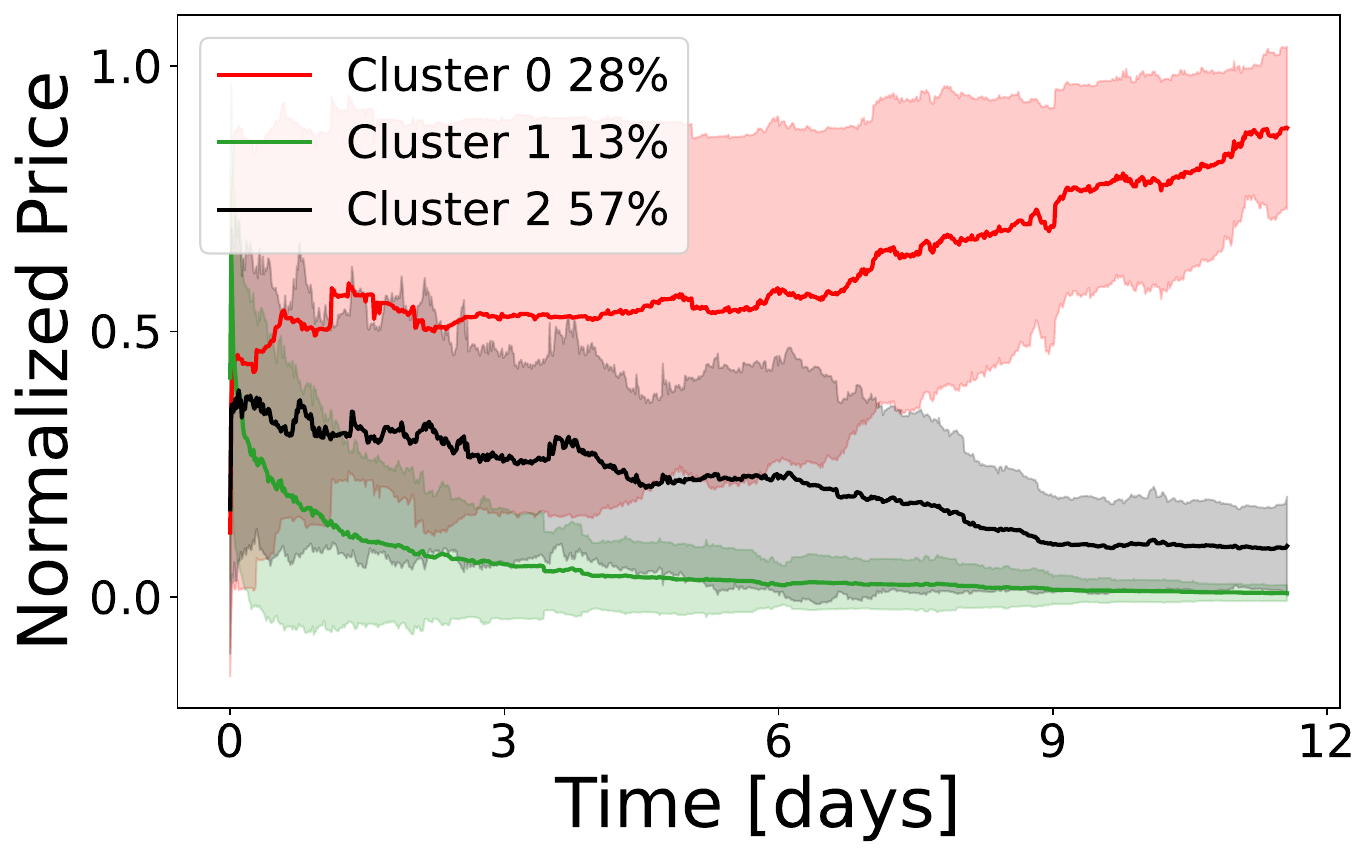}
            \caption{Clustering resulting trajectories for honeypot tokens.}
        \end{subfigure}
        \hfill
        \begin{subfigure}{0.46\linewidth}
        \includegraphics[width=8cm]{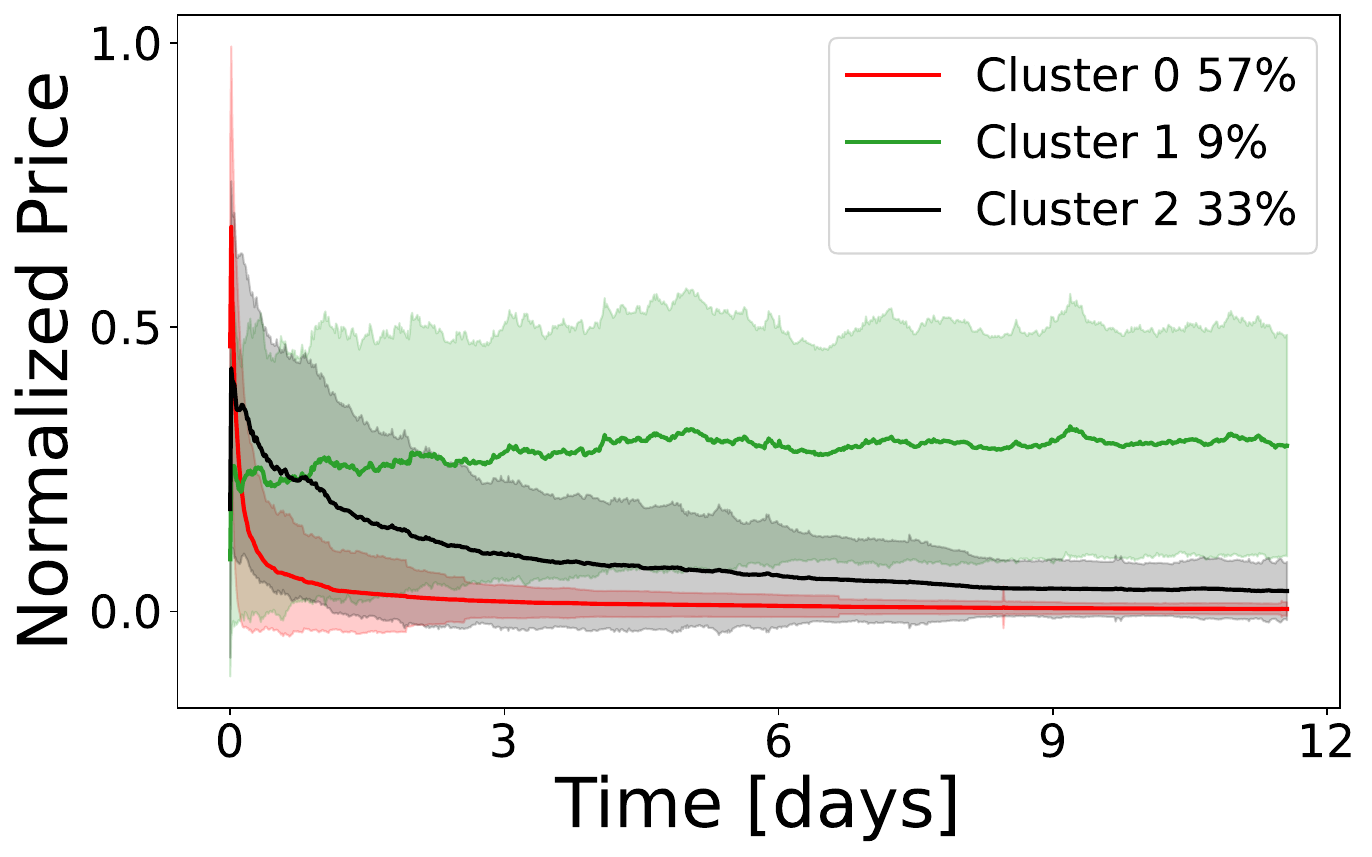}    
        \caption{Clustering resulting trajectories for sellable tokens.}
        \end{subfigure}
        \caption{The figures show the mean and the one standard deviation bands of the trajectories resulting from the clustering, with $k=3$ clusters {(see App. \ref{app:cluster} for the elbow curve analysis)}. We show the normalized price in function of the time expressed in seconds.}
        \label{dtw_honey_safe}
    \end{figure}

{\bf Clustering in physical time.} Let us now consider the case in which we consider price changes in physical time. We applied a dynamic time warping (DTW) method~\cite{Sakoe} on the financial data to categorize these tokens into clusters~\cite{Cobras}. This approach allowed us to group the sellable tokens and honeypot tokens based on their temporal price evolution (see Figure~\ref{dtw_honey_safe}). By aligning and comparing the time series, the Dynamic Time Warping (DTW) facilitated the identification of distinct patterns and behaviors, enabling a deeper understanding of the dynamics within each cluster. We apply this algorithm to the time series of normalized prices. This normalization ensures that the price oscillates between 0, when it reaches its minimum value, and 1, when it attains its maximum.

Such a definition facilitates the recognition of temporal patterns identified by the warping mechanism through a temporal shift. In this framework, the time series is expressed in terms of physical time, where transactions within the same block—sharing the same timestamp—are aggregated into a single value, with the price at that timestamp taken as the last recorded price of the block. This approach improves the robustness of pattern detection, as it mitigates irregularities due to variations in trading frequency while preserving the fundamental price dynamics.

The results of applying this algorithm to honeypots and sellable tokens are shown in Fig.~\ref{dtw_honey_safe}. A key observation is that most of these tokens experience a sharp initial price surge, followed by a rapid collapse, ultimately leading to their disappearance. The behavior of the two most populated clusters 0 and 1 we observe an extremely short lifespan of the corresponding tokens, underscoring their inherent risk. A similar pattern can also be observed in the clustering based on swap time, where the first peak represents this early price escalation, though spread over a longer period. This clustering suggests that a significant portion of the profit in a buy \& hold strategy is actually tied to securing a favorable position within a populated block, typically one of the earliest. In other words, while the strategy may indicate selling after a specific transaction, executing this action at the desired position within the block is not always feasible.
This limitation arises due to the competitive nature of transaction ordering, where traders must often pay higher gas fees to ensure priority execution. As a result, the effectiveness of the strategy is not solely dependent on price movements but also on the ability to navigate transaction ordering dynamics efficiently. It is important to stress the honeypot cluster 1, in which the normalized price appears to increase continuously over time. This represents a distinct pattern observed in certain honeypot tokens, which are designed to lure unsuspecting investors with the illusion of steady and risk-free gains. By displaying a price curve that consistently increases, these tokens create the false perception of a profitable investment, encouraging traders to buy in. However, this pattern is merely a deceptive trap, as these tokens often implement restrictions on selling, preventing investors from liquidating their holdings, and ultimately leading to financial loss.


\section{Conclusions}\label{sec:Conclusions}

In this work, we carried on a study of new tokens on the Ethereum blockchain, addressing their security issues and their financial impact on Uniswap V2. With around 15 tokens created every hour, we explored the feasibility of implementing a trading strategy in this rapidly evolving market and analyzed the challenges associated with its execution. The primary motivation for investing in this sector lies in the high potential returns that these tokens can achieve within their short life cycle. To quantify their financial significance, we introduced the concept of NTV, which allows us to assess their actual economic value within the market. Investing in this market appears to be highly lucrative, as demonstrated by a simple buy \& hold strategy. However, a closer examination of where the highest profits were generated reveals some intriguing insights. First of all, the most profitable tokens were those associated with honeypots, which, by definition, prevent uninformed investors from selling their tokens once purchased. These honeypot tokens make up approximately 88\% of the total, meaning that while they appear profitable on paper, they trap investors, making it impossible to realize gains.\\
The remaining sellable tokens--which allow for resale--still yield a positive profit, even if {to} a lesser extent. This is because the strategy assumes an immediate sell after the buy swap in the first transaction of a sandwich attack. However, in practice, executing such precise timing is impossible since most of these attacks utilize FlashBots, bypassing the public mempool and preventing external traders from positioning their transactions accordingly.We show that the high prevalence of sandwich attacks in newly launched cryptocurrencies is driven by their greater profitability in low-liquidity pools. To quantify the maximum extractable profit (excluding gas fees) from a sandwich attack, we derive an expression showing that the attacker profit is approximately equal to the victim trade size, minus adjustments that depend on the pool fee  and the reserve size  of WETH within the pair pool. In fact, deep liquidity pools mitigate sandwich attacks, as they minimize the attacker ability to move prices.\\
The presence of sandwich attacks stresses the need to consider a trading strategy based on physical time rather than swap time. This distinction becomes evident when analyzing clustering patterns in both swap time and physical time. In swap-time clustering, price movements appear smoother and less dependent on transaction ordering. In physical-time clustering, distinct clusters emerge where price paths consistently increase in the first blocks. The most populated cluster in physical time shows that for the majority of tokens the price increases in the first block and it is immediately followed by a sharp decline, leading to the token death \cite{feder2018rise}. Investing in the first blocks of a new crypto is extremely difficult because we lack control over the precise position of our transaction within the block. MEV strategies, priority gas auctions, and sandwich attacks make it nearly impossible for retail traders to execute at optimal prices.\\
{
As a first step, prospective investors approaching newly deployed tokens should rely on publicly accessible security platforms such as honeypot.is or GoPlus, which are free and provide preliminary risk assessments. However, these tools may not always offer complete coverage, particularly during the first few seconds of a token lifecycle. Certain quantitative properties can raise immediate red flags of potential honeypot behavior. For example, a small number of swaps involving unusually high amounts of WETH may indicate an artificial price increase designed to attract buyers. Similarly, a high NTV value combined with low liquidity often signals a possible rug pull, while a high NTV alone may suggest market manipulation. Most of this information is readily available through online token-scanning platforms, where the NTV can be seen as a surrogate for the market capitalization commonly displayed on such websites. For more in-depth and real-time assessments immediately after a token creation, investors may also rely on paid security services, which act as a safer bridge between uninformed users and professional researchers and/or smart contract developers.\\
}
Moreover, our study opens several avenues for future research, particularly in the areas of rug pulls, liquidity analysis, and early-block transaction dynamics. Future work should focus on developing more comprehensive methodologies to identify rug pulls at an earlier stage, potentially by analyzing smart contract characteristics before they are executed.\\

\section*{Abbreviations}
\begin{itemize}
    \item DEX =  Decentralized EXchange;
    \item V2 = Version 2;
    \item AMM = Automated Market Maker;
    \item ETH = Ethereum;
    \item WETH = Wrapped Ethereum;
    \item NTV = Net Traded Value;
    \item DTW = Dynamic Time Warping;
    \item MEV = Maximal Extractable Value;
\end{itemize}

\section*{Declarations}
\subsection*{1. Ethics approval and consent to participate}
Not applicable.

\subsection*{2. Consent for publication}
The authors consent to the publication of their article in the journal \emph{EPJ Data Science}.

\subsection*{3. Availability of data and material}
The data are publicly available through interaction with the Ethereum blockchain.

\subsection*{4. Competing interests}
The authors declare no competing interests.

\subsection*{5. Funding}
This research received funding from the following grants:
\begin{itemize}
    \item Horizon 2020 Program under the scheme “INFRAIA-01-2018-2019 – Integrating Activities for Advanced Communities”, Grant Agreement n.871042, ``SoBigData++: European Integrated Infrastructure for Social Mining and Big Data Analytics” and HORIZON Europe Program under the scheme ``INFRA-2021-DEV-02- SoBigData RI Preparatory Phase Project" Grant Agreement n. 101079043 — SoBigData RI PPP;
    \item PRIN2022 DD N. 104 of February 2, 2022 ”Liquidity and systemic risks in centralized and decentralized markets”, codice proposta 20227TCX5W - CUP J53D23004130006 funded by the European Union NextGenerationEU through the Piano Nazionale di Ripresa e Resilienza (PNRR).
\end{itemize}

\subsection*{6. Authors' contributions}
The authors contributed equally to this work.

\subsection*{7. Acknowledgements}
    FL and MN acknowledge support by the Horizon 2020 Program under the scheme “INFRAIA-01-2018-2019 – Integrating Activities for Advanced Communities”, Grant Agreement n.871042, ``SoBigData++: European Integrated Infrastructure for Social Mining and Big Data Analytics” and HORIZON Europe Program under the scheme ``INFRA-2021-DEV-02- SoBigData RI Preparatory Phase Project" Grant Agreement n. 101079043 — SoBigData RI PPP. FL and FT acknowledge support from the grant PRIN2022 DD N. 104 of February 2, 2022 ”Liquidity and systemic risks in centralized and decentralized markets”, codice proposta 20227TCX5W - CUP J53D23004130006 funded by the European Union NextGenerationEU through the Piano Nazionale di Ripresa e Resilienza (PNRR).\\

\bibliography{Bibliografia}

\begin{thebibliography}{10}
\bibitem{honeypot_url}
\bibinfo{title}{Scams, rugpulls and honeypots. what are they, and how to avoid
  them}.
\end{thebibliography}

\newpage
\appendix

\section*{Appendix}
\addcontentsline{toc}{section}{Appendix}

\section{New token{s} Created: Additional Plots}
\label{app_num_coins}

\begin{figure}[H]
\centering
\begin{subfigure}{0.46\textwidth}
  \centering
  \includegraphics[width=\linewidth]{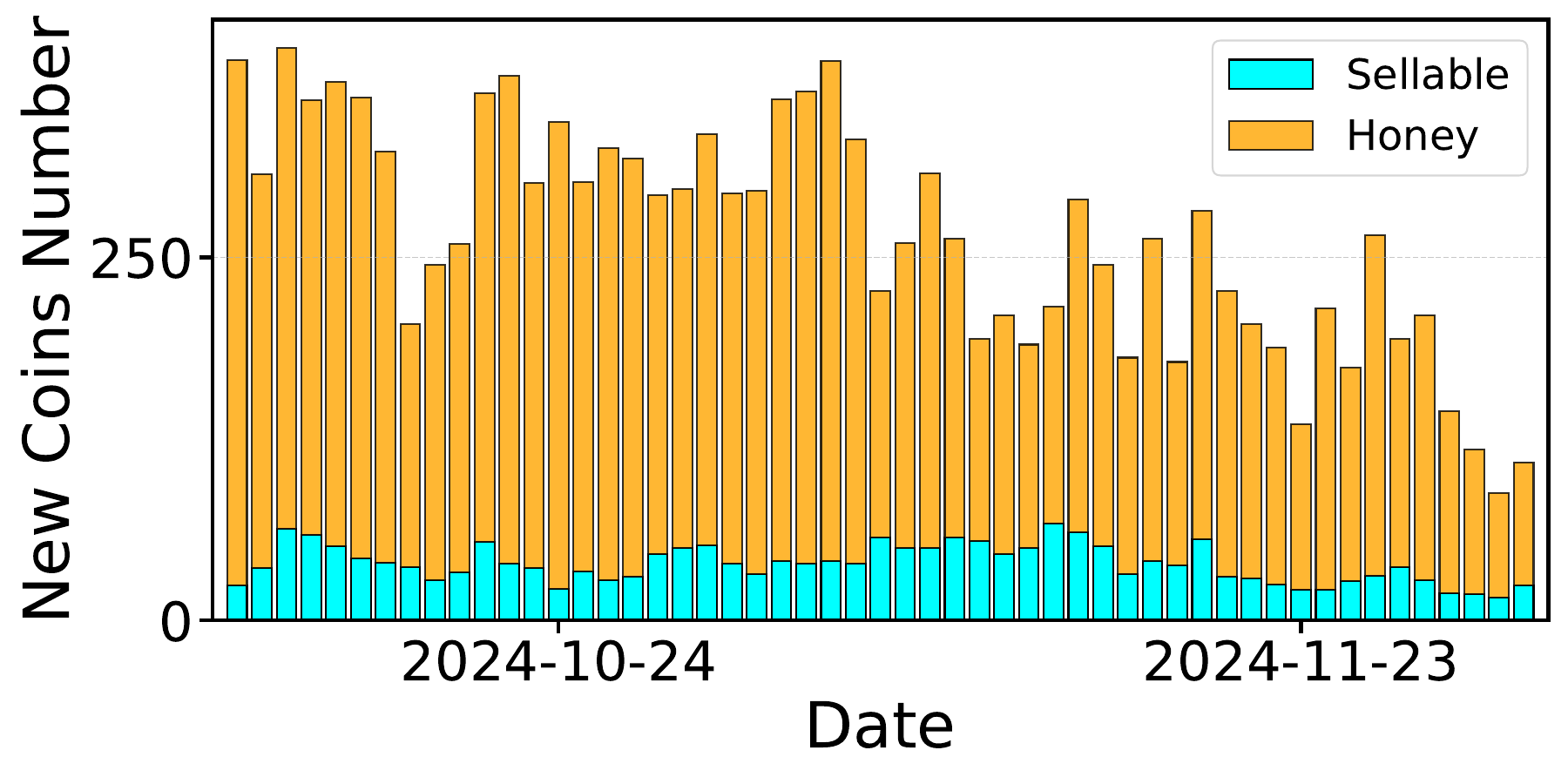}
  \caption{Number of new tokens created daily during the period covered by our dataset.}
  \label{fig:Daily_Creation}
\end{subfigure}
\hfill
\begin{subfigure}{0.46\textwidth}
  \centering
  \includegraphics[width=\linewidth]{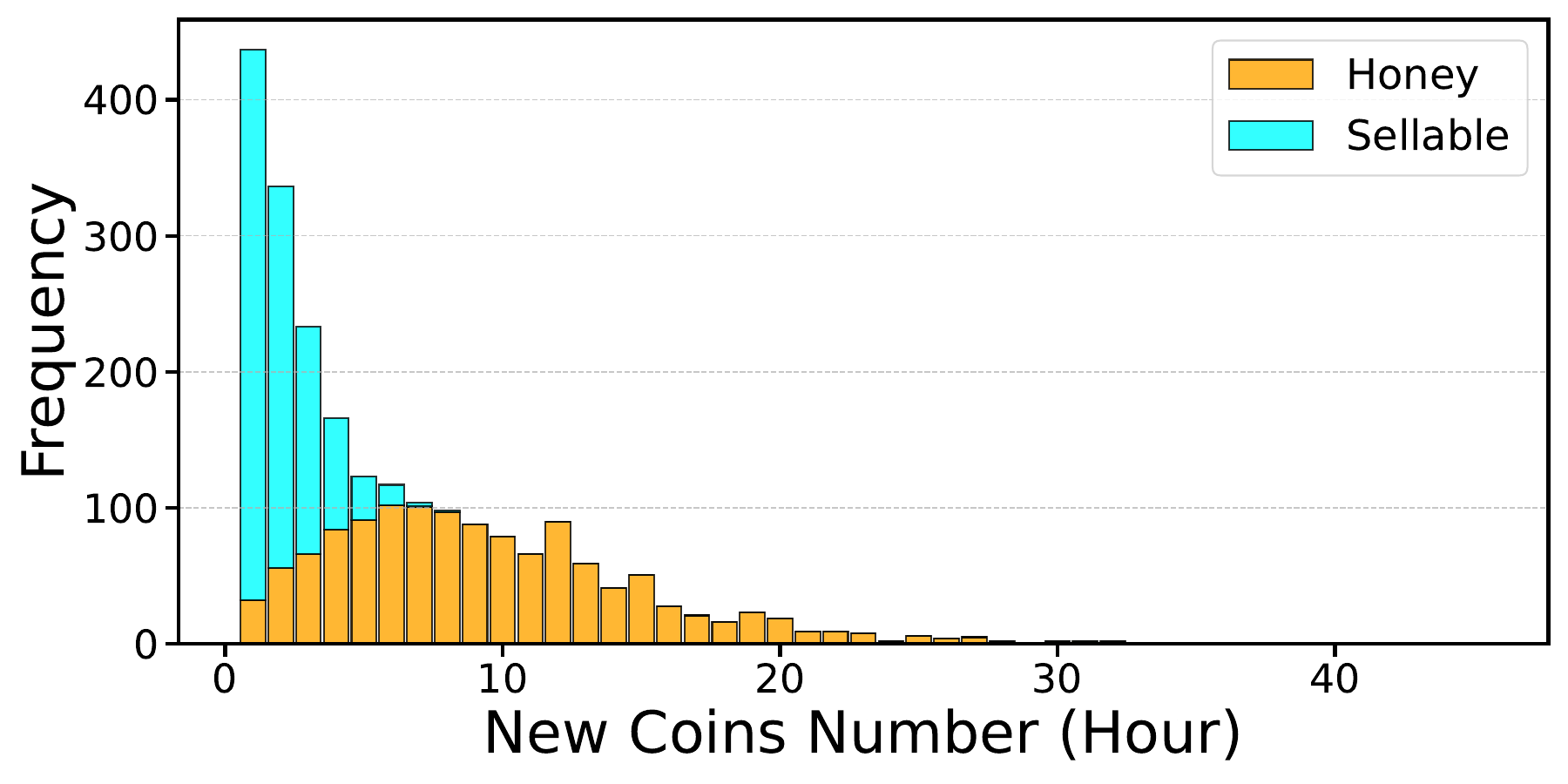}
  \caption{Distribution of the number of new tokens created per hour. $\qquad$}
  \label{fig:BinnedHour}
\end{subfigure}
\caption{{This visualization uses stacked bars.}}
\label{fig:corr_am}
\end{figure}

In the Figure~\ref{fig:Daily_Creation}, we presents a histogram displaying the number of tokens created each day, while Figure~\ref{fig:BinnedHour} illustrates the distribution of the number of tokens created within one-hour bins.

\section{{Return performances after the creation.}}\label{App:Returns}

One can examine the distribution of returns for the new coins at $N_t$ swaps after their creation, analyzing the variation in price from their inception to that point. This can provide insight into their trends during the early stages of their existence. We define the log returns as 
\begin{equation}
    r^i_{N_t}(0) = \log\bigg(\frac{p^i({N_t})}{p^i(0)}\bigg),
\end{equation}
where $p(0)$ is the initial price at which the token is created. Here, $p^i(t=N_t)$ is the token price recorded at the $N_t$-th swap for the $i$-th token. We take $N_t = \{10, 100,1000\}$.

\begin{figure}[H]
\centering
\begin{subfigure}{0.46\textwidth}
\includegraphics[width=\linewidth]{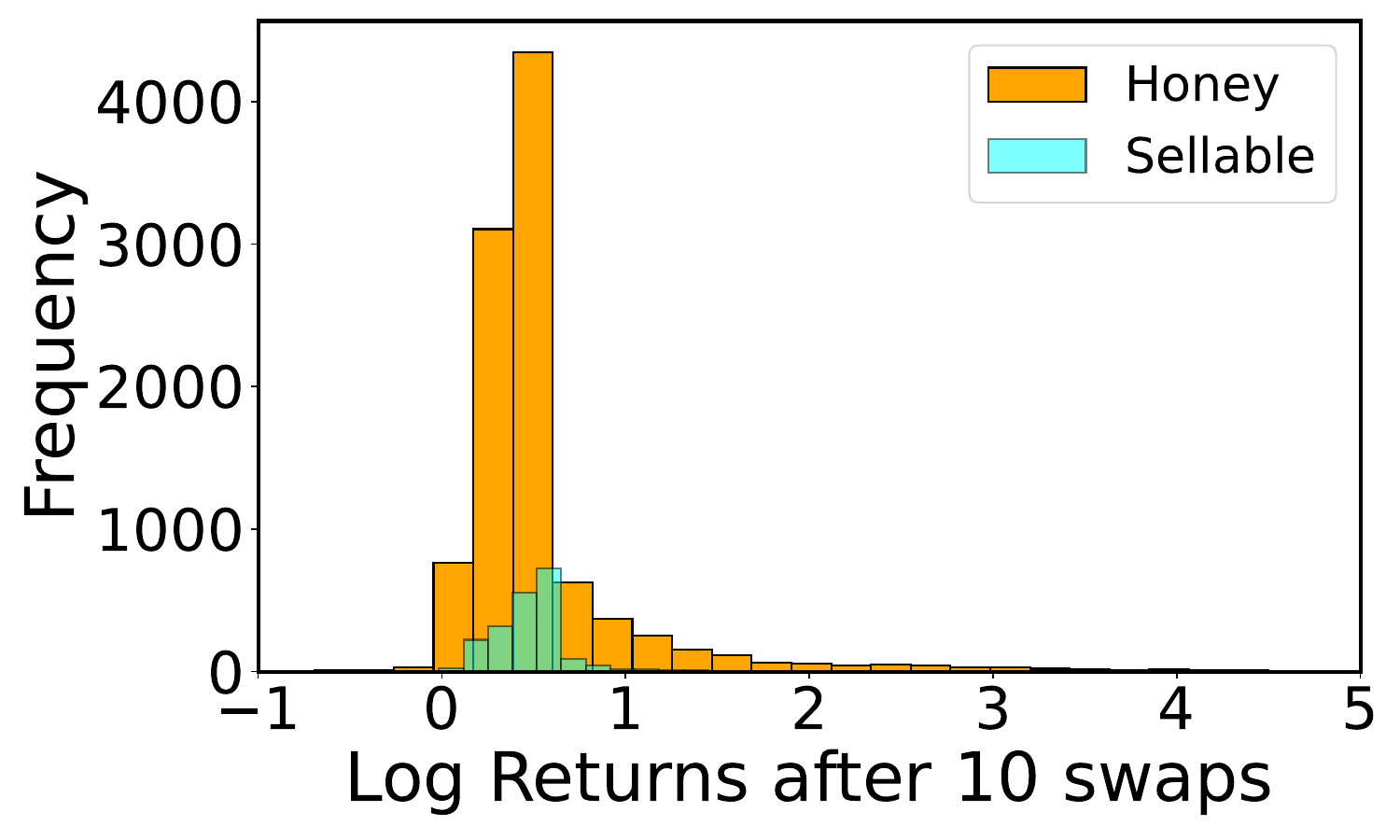}
  \caption{Histogram of lagged log returns after 10 swaps. The minimum value for sellable tokens is -2.41 {W}ETH while the maximum is 10.90. Instead for honeypots are respectively -30.56 and 34.37.}
   \label{fig:Histo_Cap_10}
\end{subfigure}
\begin{subfigure}{0.46\textwidth}
  \centering
\includegraphics[width=\linewidth]{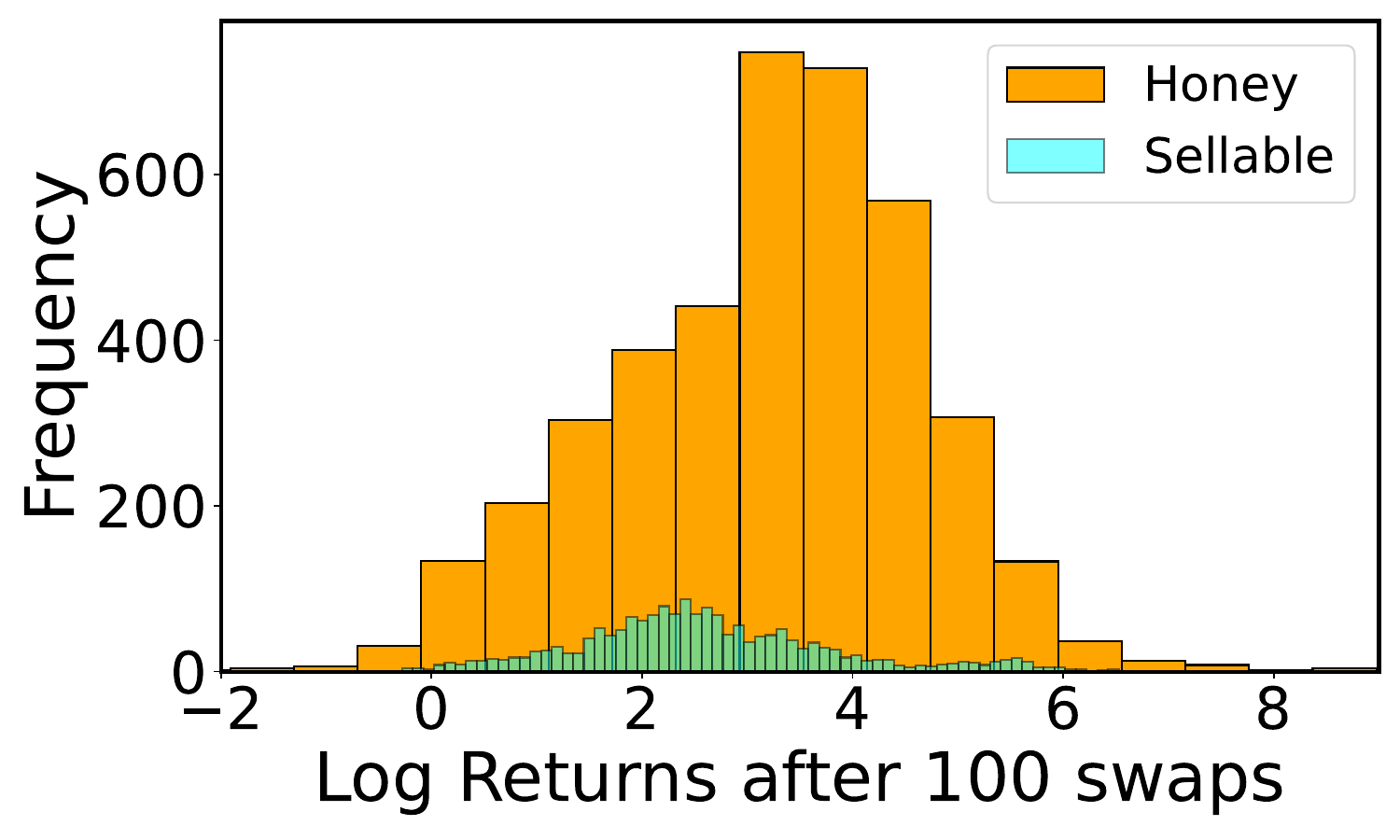}
  \caption{Histogram of lagged log returns after 100 swaps. The minimum value for sellable tokens is -2.82 while the maximum is 7.34. Instead for honeypots are respectively  -30.94 and 29.54.} 
  \label{fig:Histo_Cap_100}
\end{subfigure}
\caption{Histograms of the lagged log returns of the new tokens after different number of swaps.}
\label{fig:Istogrammi_Ret}
\end{figure}

\begin{figure}[H]
  \centering
\includegraphics[width=8cm]{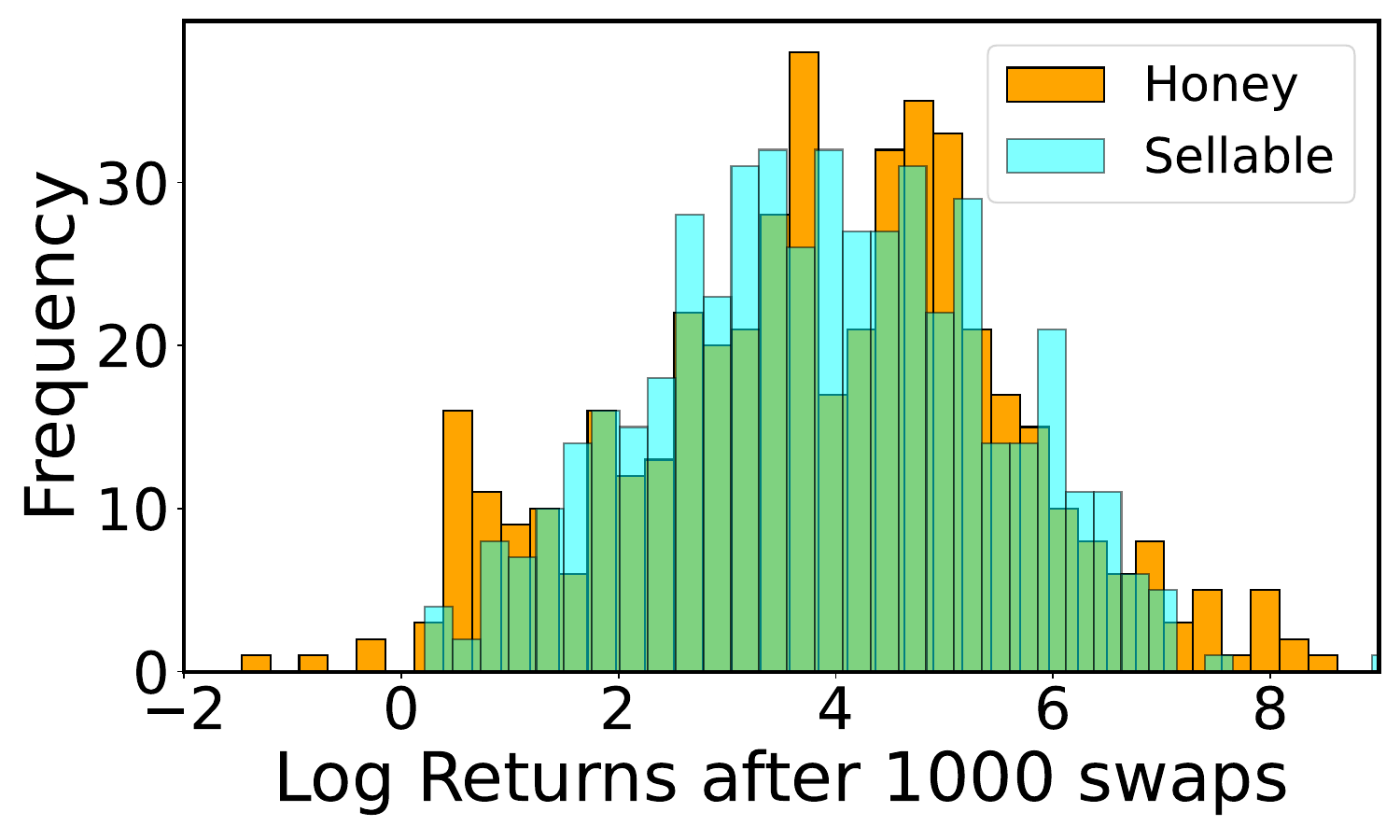}
  \caption{Histogram of lagged log returns after 1000 swaps. The minimum value for sellable tokens is -3.63 while the maximum is 9.19. Instead for honeypots are respectively  -1.47 and 11.80.}
  \label{fig:Histo_Cap_1000}
\end{figure}
In Figure~\ref{fig:Istogrammi_Ret} we show the histograms of the lagged returns for the different values of $N_t$. The distribution of the lagged returns is in great percentage concentrated on positive values both for sellable and honeypots tokens. This indicates that if the token maintains a non-zero order flow, its value with high probability increases. This is reasonable also considering that these tokens are often created on very small values, e.g. $\sim 10^{-8}$ WETH. Also in this case, as expected, extreme values for the returns are present, particularly for the honeypots tokens which, however, usually become inactive very soon.
\begin{figure}[H]
  \centering
\includegraphics[width=0.46\textwidth]{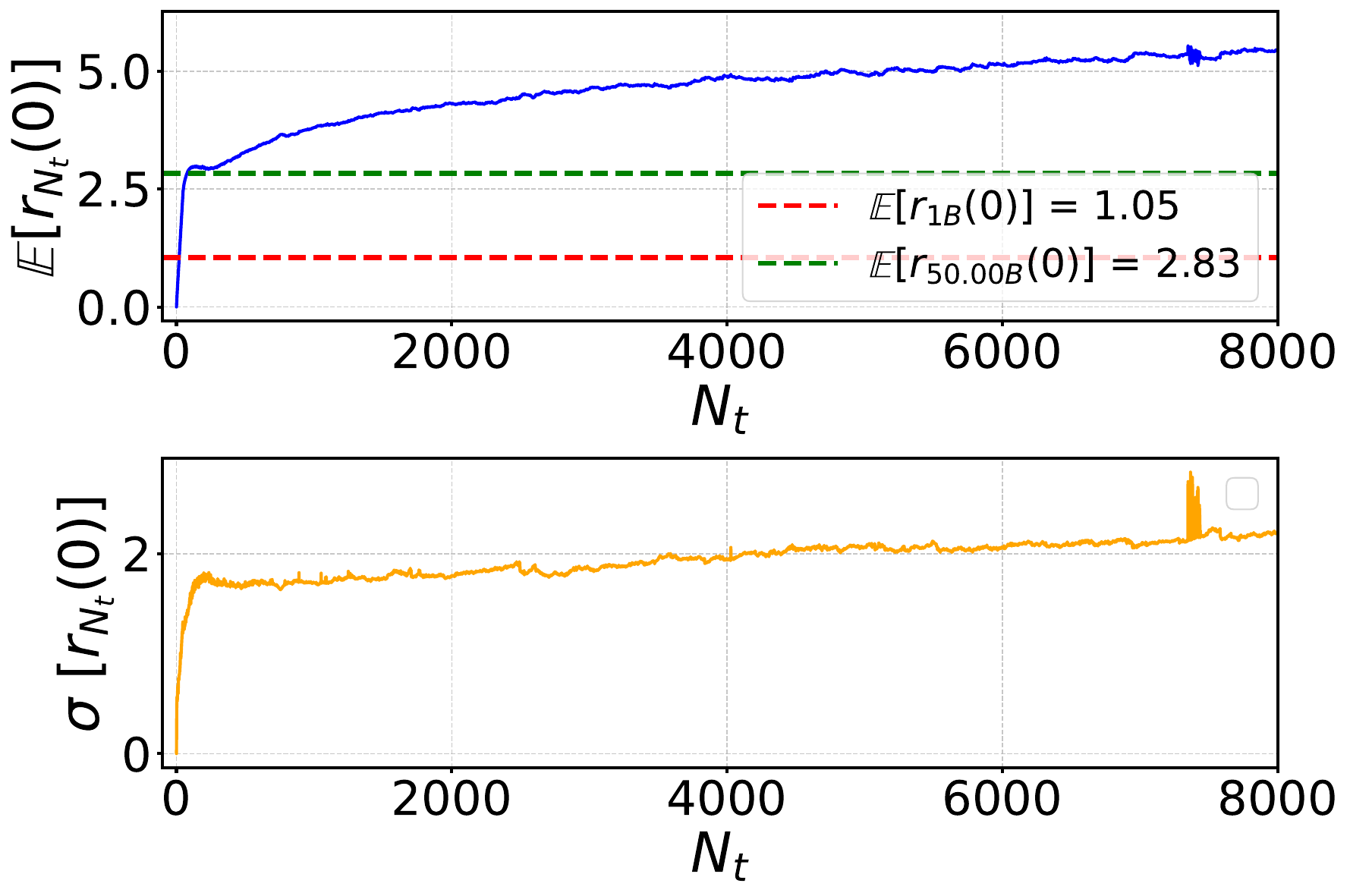}
  \caption{Expectation value and standard deviation of $r_{N_t}(0)$.}
  \label{fig:EV_r0}
\end{figure}
Given these returns, even retail investors with limited capital can acquire a significant amount of tokens and potentially generate substantial profits if they invest at the very early stages of a token's lifecycle. However, this is only true if the investment is made in a sellable token. If the token turns out to be a honeypot or is subject to manipulative practices such as a rug pull, the likelihood of realizing any gains becomes negligible.
Now, an important question is what is the expectation value and the variance of $r_{N_t}(0)$ as a function of $N_t$.  Thus, if we suppose to be able to buy the coin at $t=0$, which is the average return and the variance on it after $N_t$ steps. In Fig.~\ref{fig:EV_r0}, we show that after a rapid increase they both tend to decelerate. The first increase of the returns is, on average, due to the first 50 blocks actions. The first block contains on average $\sim 27$ swaps. Then, the number of swaps tends to decrease. This, together with the fact that the number of coins decreases as the number of swap increase, is probably the reason why the first rapid increase tends to decelerate. Indeed in the limit of large swap the remaining coins tend to stabilize. 
\begin{figure}[H]
  \centering
\includegraphics[width=0.46\textwidth]{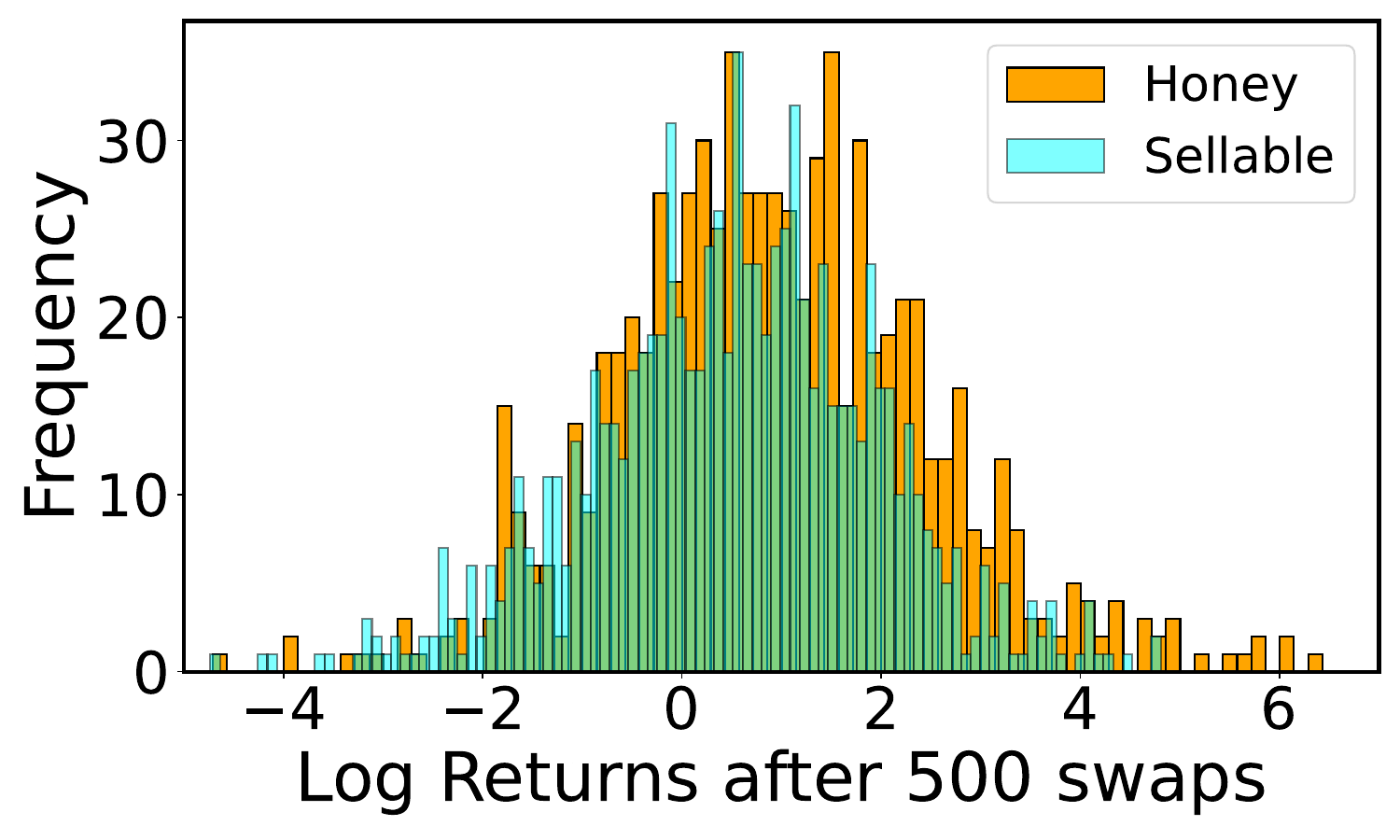}
  \caption{The figure shows the distribution of $r_{500}(60)$.}
  \label{fig:LogRetIn60}
\end{figure}
\begin{figure}[!htb]
  \centering
\includegraphics[width=0.46\textwidth]{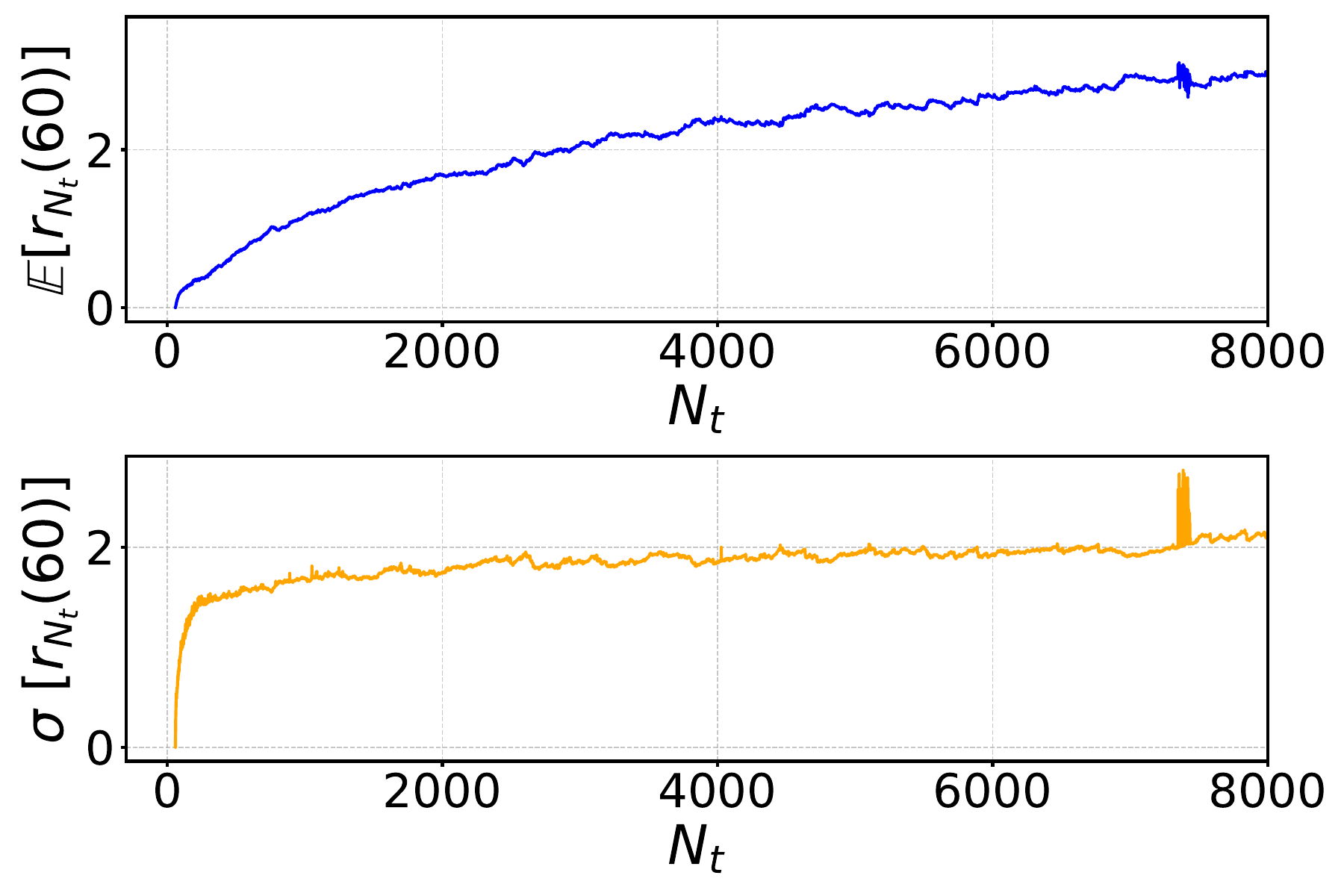}
  \caption{Expectation value and standard deviation of $r_{N_t}$(60) as a function of $N_t$. It is interesting to observe that the expectation value keeps increasing while the standard deviation saturates.}
  \label{fig:EV_In60}
\end{figure}
For inexperienced investors who are unable to distinguish between honeypots and legitimate projects, the new coins market in Uniswap v2 can easily appear to be a goldmine. Indeed, this analysis shows that the lagged log returns, measured from the coin's creation, are predominantly positive. This perception likely contributes to the sustained participation of investors in these coins. 
As mentioned earlier, the first block typically contains around 27 swaps on average, and this number decreases as the swap count increases.
During the initial blocks, gas fees are extremely high, as there is naturally a bidding race to be among the first to enter.
Moreover, participating in these early blocks requires a certain degree of technical sophistication.
It is therefore unrealistic to assume that the returns just presented could be achieved by an average investor.
For this reason, we repeat the same analysis by taking the 60th swap as the starting point.
Although this choice is somewhat arbitrary, it allows a realistic number of blocks to elapse, making entry feasible even for less sophisticated investors. 
As we show, similar performance levels are nevertheless obtained under this assumption. Furthermore, small variations around this threshold do not lead to significant changes in the results, as returns remain very high overall. In Figure~\ref{fig:LogRetIn60}, we present the distribution of $r^{500}(60)$ for the subset of tokens exceeding this threshold. The distribution spans both negative and positive values, with the mean and median—indicated by dashed lines—both lying in the positive region. This suggests a general tendency toward positive log returns for active tokens, a trend further corroborated by Figure~\ref{fig:EV_In60}.

\section{Sandwich Attack}
\label{appendix:sandwichAttack}

\begin{figure}[H]
\centering
\begin{subfigure}{0.46\textwidth}
  \centering
  \includegraphics[width=\linewidth]{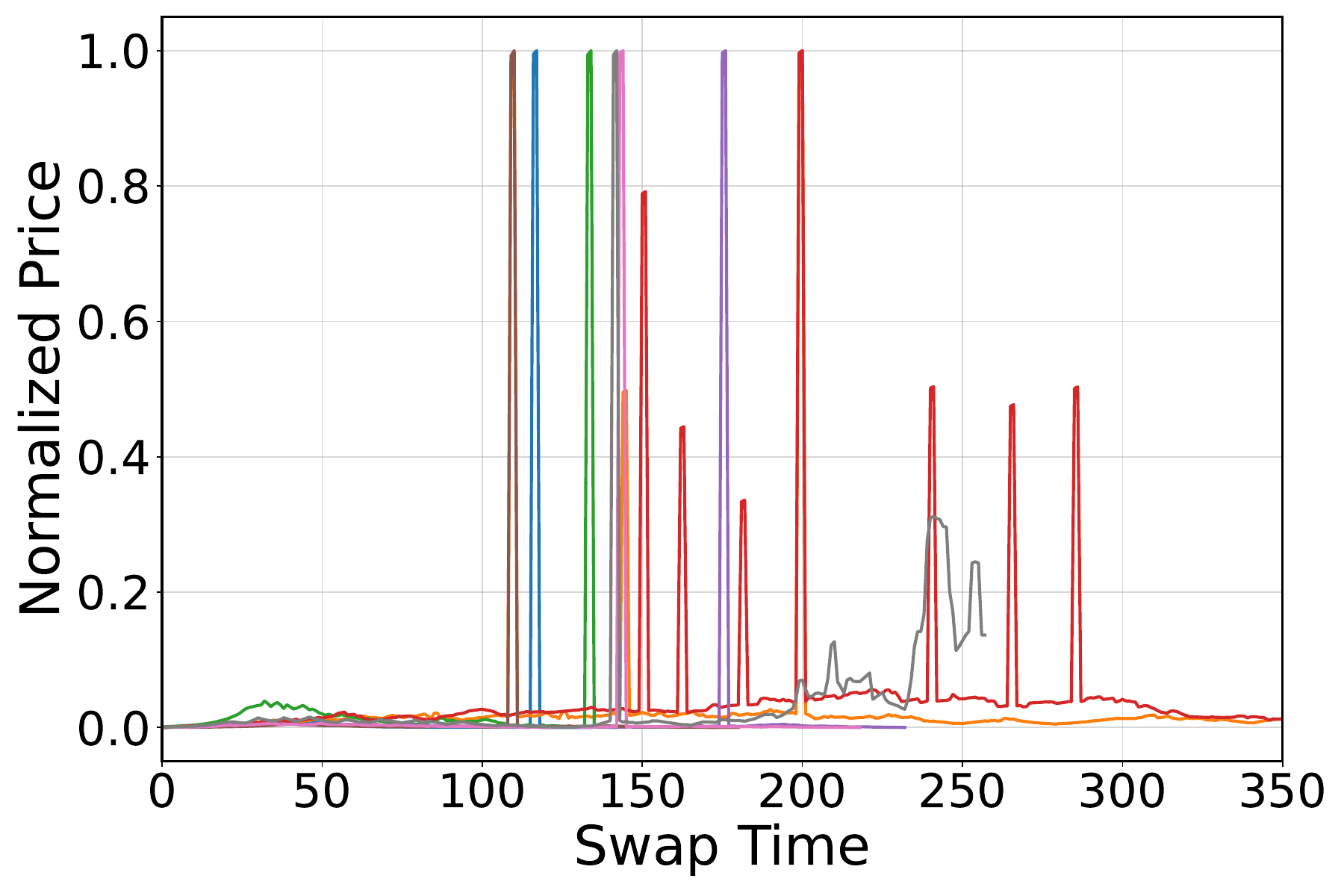}
\end{subfigure}
\begin{subfigure}{0.46\textwidth}
  \centering
  \includegraphics[width=\linewidth]{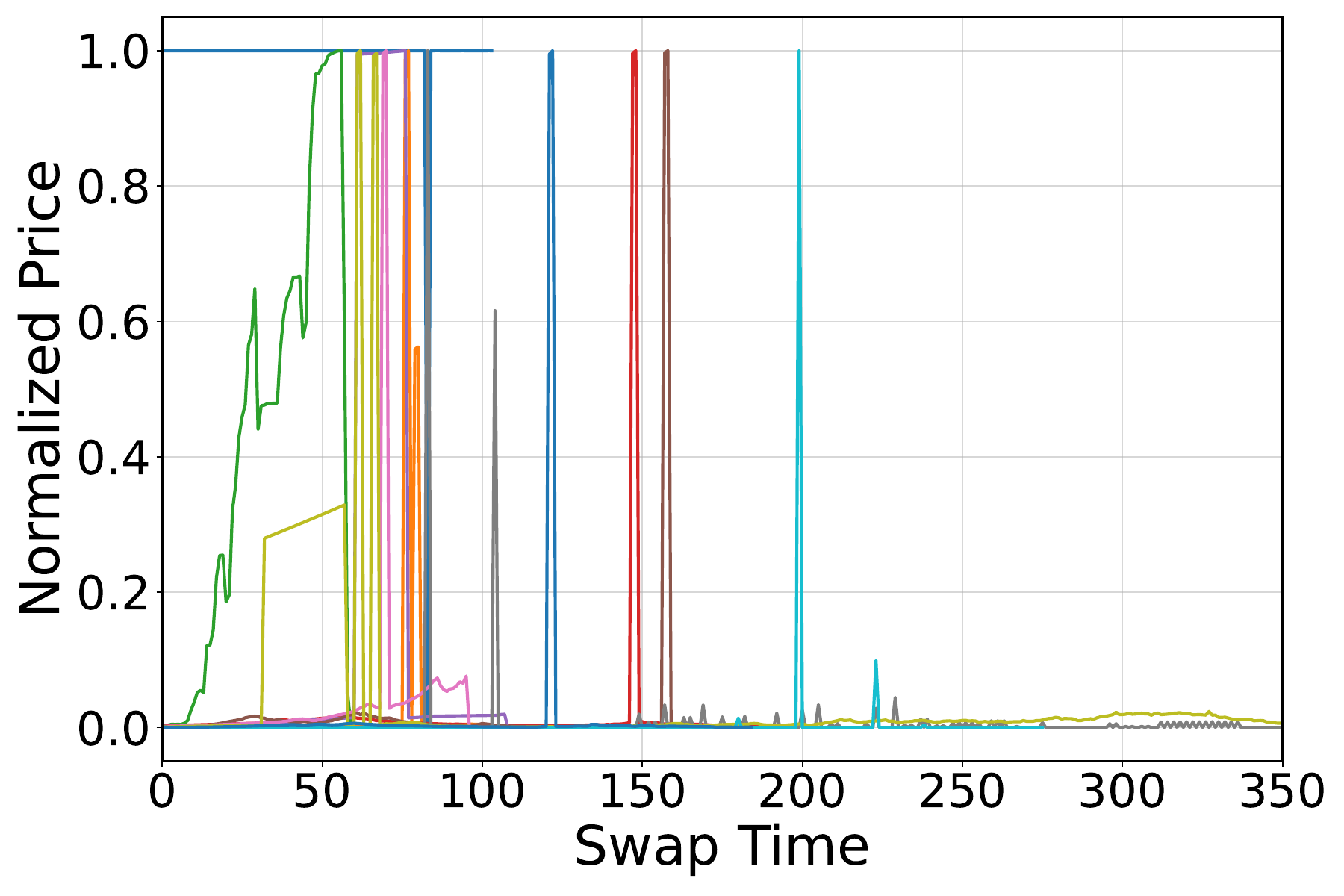}
\end{subfigure}
\caption{The figures show the normalized price trends using a min-max scaler for the first 350 swaps. The right panel displays all sellable coins that generated a profit greater than 0.5 {W}ETH, while the left panel shows the same for honeypots. Sharp peaks reaching large values and immediately dropping can be observed for all coins, corresponding to sandwich attack events. There are only buy swap sandwiches because at the end of the three transactions, the attacker remains with only WETH.}
\label{fig:Price_Peaks_Figures}
\end{figure}

Figure~\ref{fig:Price_Peaks_Figures} illustrates the normalized price evolution over the first 350 swaps for all coins that yielded a profit greater than 0.5 {W}ETH in our strategy. Sharp price spikes followed by immediate drops correspond to sandwich attacks, impacting both sellable tokens and honeypots. Our strategy capitalizes on these peaks, securing profits before prices revert. In the absence of such attacks, the strategy would generate negative returns and would therefore be inapplicable for achieving positive profits in real investments.

Table~\ref{table:sandwich} provides an example where a front-runner gains $s = 0.0316$ {W}ETH from a $25$ {W}ETH investment.
\begin{table}[!htb]
\centering
\begin{adjustbox}{max width=\textwidth}
\begin{tabular}{l|l|l|l|l|l|l}
\toprule
\textbf{$\Delta x$} & \textbf{$\Delta y$} & \textbf{$x_{\rm new}$} & \textbf{$y_{\rm new}$} & \textbf{New Price} & \textbf{Old Price} & \textbf{Liquidity} \\
\hline
25.0000 & -268464000 & 28.3199 & 35758537 & 7.920e-07 & 1.091e-08 & 31823 \\
0.0500 & -62833 & 28.3699 & 35695704 & 7.948e-07 & 7.920e-07 & 31823 \\
-25.0316 & 268463996 & 3.3382 & 304159700 & 1.098e-08 & 7.948e-07 & 31865 \\
\bottomrule
\end{tabular}
\end{adjustbox}
\caption{Sandwich attack on the token AIDLE (pair address $=$ "0xdab46ad9dc16d3de41fdcdc4eabd26fe8b188d91") executed in the block 21560868. Here, New Price indicates the price after the swap, while Old Price the price before.}
\label{table:sandwich}
\end{table}
This mechanism executed by the attacker does not correspond to the optimal execution of the sandwich attack.

Detecting pending transactions in the mempool of the Ethereum blockchain allows traders to anticipate the intentions of others before their transactions are executed. Specifically, by monitoring transactions targeting swap router contract addresses, it is possible to identify when a trader is about to buy a given quantity of a token, whose reserve is denoted with \( y \), in exchange for a certain amount of Ethereum, whose reserve is \( x \). 

\subsection{Optimal Strategy}

To analyze the optimal strategy in detail, we consider the execution order of the three key transactions involved in a sandwich attack:\\
- The {\bf first swap} by the sandwich attacker, who buys a token amount \( \Delta y_a \) and sells \( \Delta x_a \) Ethereum. Using the constant product formula $( x + r\Delta x_a ) (y - \Delta y_a) = x y$ :
$$
\Delta y_a = \frac{ r \Delta x_a  y }{ x+r\Delta x_a  } \,\,;
$$
where $r = 1 - f$ $(f=0.3\% \ll 1)$.\\
- The {\bf swap of the victim}, who buys a token amount \( \Delta y _\epsilon \) and sells \( \Delta x_\epsilon \):
$$
\Delta y _\epsilon = \frac{ r \Delta x_\epsilon (y - \frac{r\Delta x_ay}{x+r\Delta x_a}) }{ x+\Delta x_a + r \Delta x_\epsilon } \,\,;
$$
- The {\bf second swap} of the attacker who sells all of his previously purchased bought tokens $\Delta y_2 = \frac{ r\Delta x_ay }{ x+r\Delta x_a }$ and buy $\Delta x_{tot}$ Ethereum:
$$
\Delta x_{tot} = r^2 \Delta x_a  \frac{ (x+\Delta x_a+\Delta x_\epsilon) (x+\Delta x_a+r \Delta x_\epsilon) }{ x^2  + r^2 \Delta x_a (x+\Delta x_a) + r^3 \Delta x_a \epsilon + \Delta x_ax } \,\,.
$$
We have tested the correctness of this result by running simulations using the core Uniswap V2 code. 

We define with $s$ the total amount of Ethereum that the sandwich attacker gains after all his transactions, that is, the total amount $\Delta x_{\rm tot}$ obtained in the last swap minus the first swap $\Delta x_a\,$:
$$
s = \Delta x_{tot} - \Delta x_a \,\,.
$$
Note that in the case where the value of $\Delta x_\epsilon$ is $0$ we obtain the amount $\Delta x_{\rm tot}$ characterized only by the pool fee which leads to an asymptotic loss given by
\begin{equation}\label{eq: AsymptLoss}
   s_{\Delta x_\epsilon = 0 } = \Delta x_{tot} - \Delta x_a = \frac{r^2 \Delta x_a (x+\Delta x_a)}{x + r^2 \Delta x_a} - \Delta x_a = \frac{(r^2 - 1 ) \Delta x_a x}{x + r^2 \Delta x_a}
   \xrightarrow[\Delta x_a \to \infty]{ }\Biggl( 1 - \frac{1}{r^2} \Biggl) x \sim -2fx\,\,.
\end{equation}
Thus, there is an upper limit to the loss we can have in the limit of infinite investment and it depends on the reserve $x$ and on the fee $f$. Let us now consider the complete expression of $s$
\begin{equation}\label{eq: Completes}
    s(\Delta x_a; \Delta x_\epsilon) = \frac{ \Delta x_a r^2 ( x + \Delta x_a)( x + \Delta x_\epsilon ) + \Delta x_a\Delta x_\epsilon r^3 ( x +  \Delta x_\epsilon ) - \Delta x_a x (x+\Delta x_a) }{( x + r^2 \Delta x_a) (x+\Delta x_a) + r ^3 \Delta x_a \Delta x_\epsilon } \,\,.
\end{equation}
 In the absence of fee, i.e., when \( r = 1 \), the maximum gain \( s \) is achieved in the limit \( \Delta x_a \rightarrow \infty \) and corresponds exactly to \( \Delta x_\epsilon \). Thus, for the attacker it is convenient to invest as much as possible in the sandwich attack. When fees are present, an important trade-off arises. As the attacker increases its investment \( \Delta x_a \), the absolute profit potential initially grows tending to $\Delta x_\epsilon$, but so does the total fee paid to the liquidity pool, i.e. $f\Delta x_a$. At a certain point, the accumulated transaction fees outweigh the marginal profit gained from additional investment. This means that beyond a certain threshold, increasing \( \Delta x_a \) further leads to diminishing or even negative net returns, as the cost of executing the attack surpasses the maximum potential gain \( \Delta x_\epsilon \). Consequently, when fees are taken into account, the optimal attack size must balance maximizing price impact while minimizing excessive fee costs. For this reason we expect that maximal gain will happen in a regime where the two quantities are of the same order, i.e. $f \Delta x_a \simeq \mathcal{O}(\Delta x_\epsilon)$.

Indeed, if we consider $f \Delta x_a \ll \Delta x_\epsilon$, we practically reduce to the case in which the fee is not important, i.e. $f=0$, leading the investor to increase as more as possible its investment. This condition ends to be true when the loss due to the fee becomes to be relevant, i.e. $f \Delta x_a \simeq \mathcal{O}(\Delta x_\epsilon)$. On the other side, i.e. for $f\Delta _a \gg \Delta x_\epsilon$, the fee becomes more and more relevant leading to increasing losses. 

Thus we can guess that the maximum lies in the region where $\Delta x_\epsilon/\Delta x_a$ and $f$ are of the same order. Thus we can write
$$
s = \Delta x_a \Biggr[\frac{ \Delta x_\epsilon/\Delta x_a } { 1 + \frac{ x^2 }{ \Delta x_a(2x+\Delta x_a) }  } - 2 \frac{ x }{ x+\Delta x_a } f \Biggr] +  O(\Delta x_\epsilon^2/\Delta x_a^2)\,\,,
$$
where, since $f\ll 1$, we can reasonably neglect the second order. In order to find the maximum gain $s_{\rm max}$, we compute its derivative respect to $\Delta x_a\,$:
\begin{align}
\frac{\partial s}{\partial(\Delta x_a)} & = 2 x^2 \frac{ \Delta x_\epsilon (x+\Delta x_a)^3 - f \biggl( \Delta x_a (2x+\Delta x_a) + x^2 \biggl)^2 }{\Bigl( \Delta x_a (2x+\Delta x_a) + x^2 \Bigl)^2 (x+\Delta x_a)^2} \,\,,  
\end{align}
the maximum is for
\begin{equation}\label{eq: MaxPoint}
\Delta x_a^{\rm max} = \frac{ \Delta x_\epsilon }{f} - x, \,\,
\end{equation}
expressing $s_{\rm max}$ using the value of $\Delta x_a^{\rm max} \,$:
\begin{equation}\label{eq: Maximum}
s_{\rm max} = \frac{ ( \Delta x_\epsilon - fx )^2 }{ \Delta x_\epsilon } \,\,,
\end{equation}

\begin{figure}[!htb]
  \centering
  \includegraphics[width=0.6\textwidth]{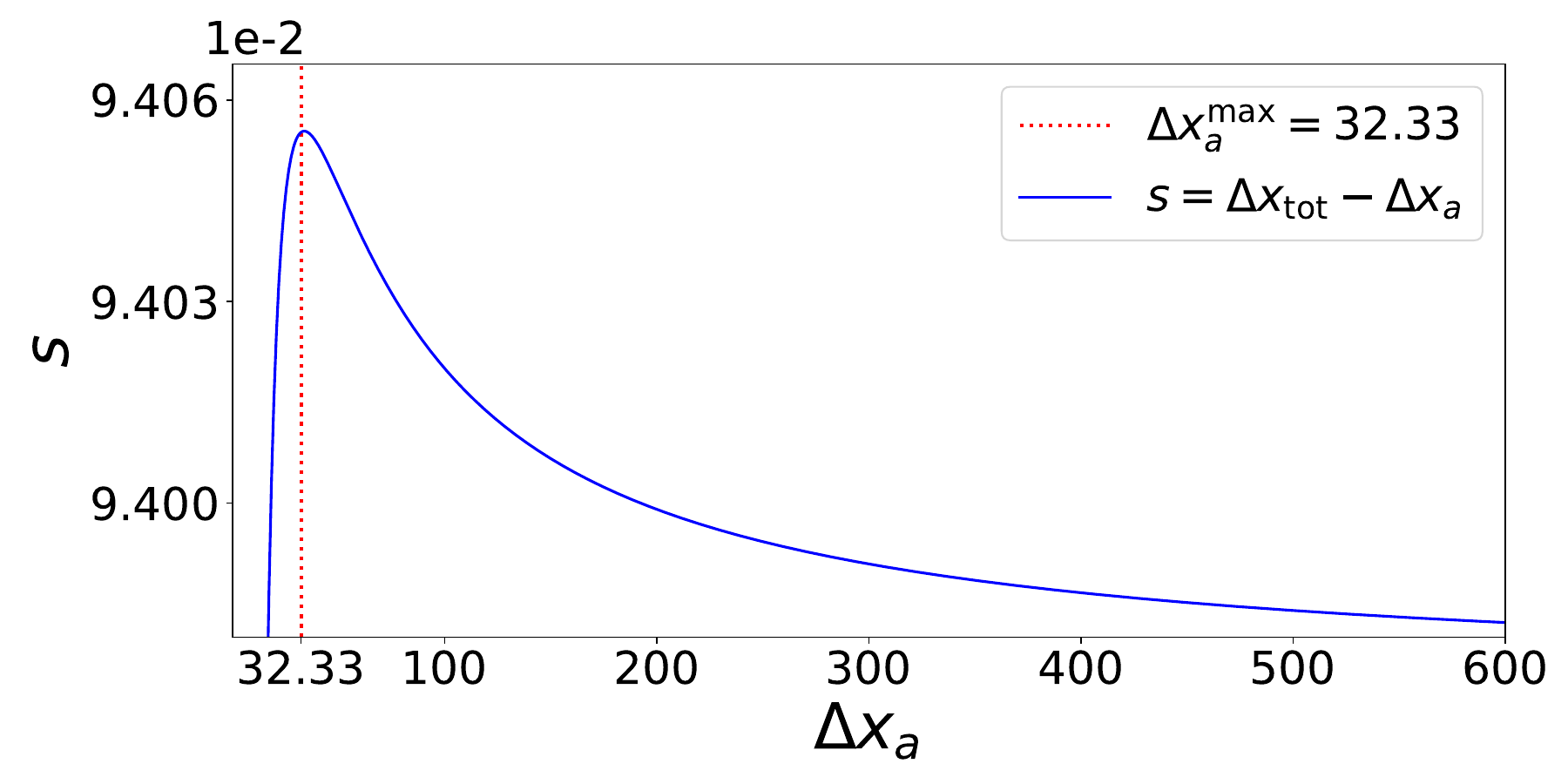}
  \caption{Visualization of the function \( s \) in Eq.\eqref{eq: Completes} as a function of the attacker input amount \( \Delta x_a \). The plot is generated using the parameter values \( x = 1 \), \( r = 0.997 \), and \( \Delta x_\epsilon = 0.1 \). The red vertical line corresponds to the estimated maximum point of Equation~\eqref{eq: MaxPoint}.}
  \label{fig:Max_Sandw}
\end{figure}

Figure~\ref{fig:Max_Sandw} shows the behavior of the complete expression for \( s \) in Eq.~\eqref{eq: Completes} as a function of the attacker input amount \(\Delta x_a\), keeping the victim amount $\Delta x_\epsilon$ fixed. The observed trend shows a concave increase up to a peak, followed by a decrease that eventually approaches a plateau for large values of \(\Delta x_a\). The initial rise in profit is driven by the increasing size of the attacker investment. This growth is rapid until the portion of the investment lost to the pool fee becomes significant enough to counterbalance the gain. We observe that the estimated maximum $s_{\rm max}$ is located very close to the point predicted by our approximation.
Beyond this peak, as the pool fee becomes more dominant, the gain decreases asymptotically, as predicted by Equation~\eqref{eq: AsymptLoss}. It is important to note that the presence of a nonzero \(\Delta x_\epsilon\) ensures a positive contribution to the gain, preventing it from turning negative. Specifically, for \(\Delta x_\epsilon \neq 0\), we obtain:
\begin{equation}
    s(\Delta x_a;\Delta x_\epsilon) \rightarrow \frac{r^2(x+\Delta x_\epsilon)-x}{r^2}, \quad \text{for } \Delta x_a \rightarrow \infty.
\end{equation}
For \(\Delta x_\epsilon = 0\), we recover the result previously derived in Equation~\eqref{eq: AsymptLoss}, while for \( r = 1 \), the maximum gain simplifies to \(\Delta x_\epsilon\). 

{
\section{Clustering Informativeness and Robustness Analysis}
\label{app:cluster}
In this section, we present a detailed study of the clustering structure and the robustness of our results. 
We start by examining the case in \textit{swap time}, which captures the intrinsic trading dynamics independently of physical time aggregation. 
To evaluate how informative the chosen number of clusters is, we first report in Fig. \ref{fig:elbow} the \textit{elbow curve}, obtained by plotting the within-cluster sum of squares (WCSS) as a function of the number of clusters $k$.
\begin{figure}[H]
    \centering
    \includegraphics[width=8cm]{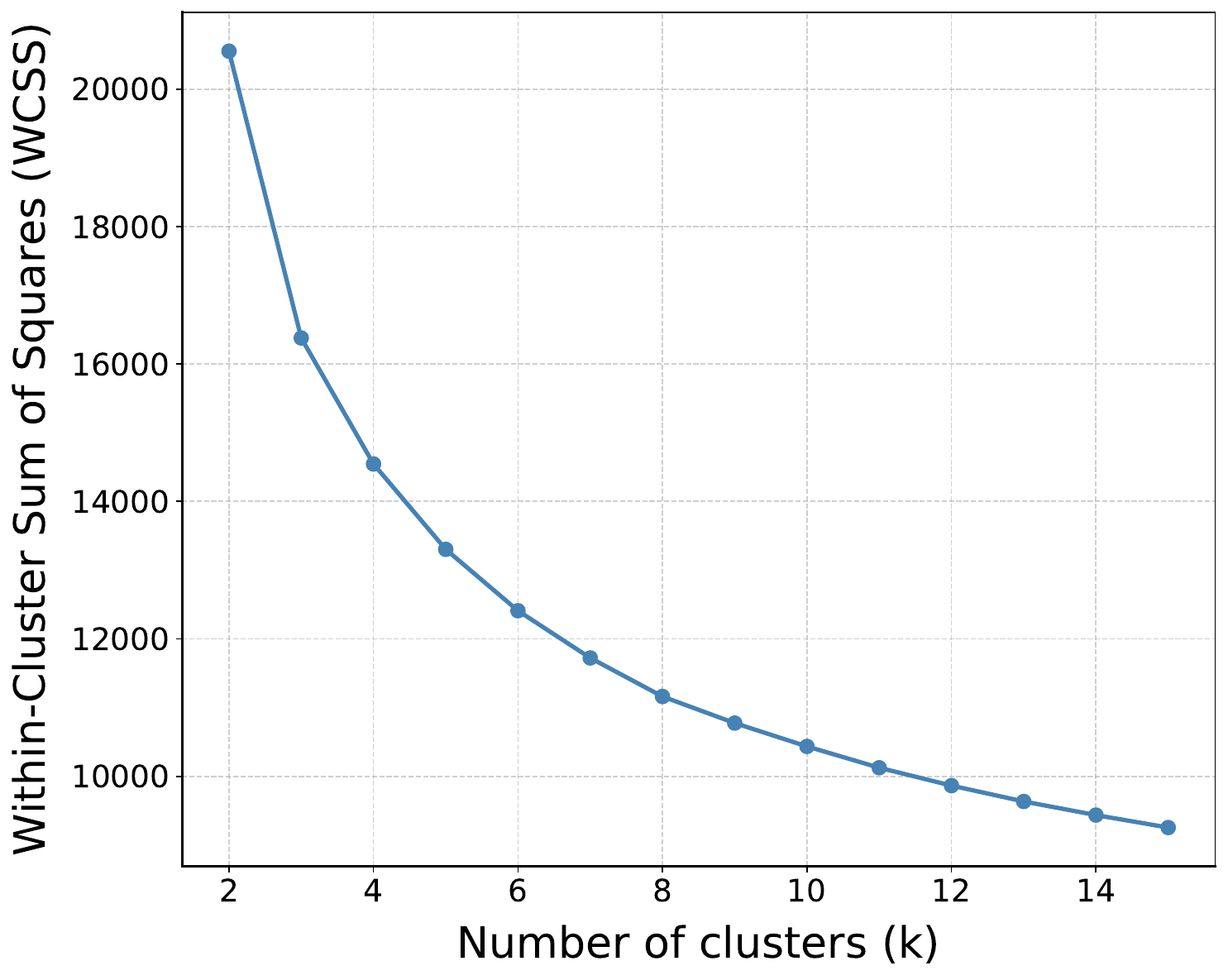}
    \caption{\textbf{Elbow curve.} Within-cluster sum of squares (WCSS) as a function of the number of clusters $k$. 
    The inflection point between $k=2$ and $k=4$ indicates the region where additional clusters yield diminishing improvements in compactness.}
    \label{fig:elbow}
\end{figure}
The curve shows a clear inflection point between $k=2$ and $k=4$, suggesting that the main structure of the data is effectively captured within this range. 
We then visualize the resulting partitions through principal component analysis (PCA). 
\begin{figure}[H]
    \centering
    \begin{subfigure}{0.46\linewidth}
        \includegraphics[width=8cm]{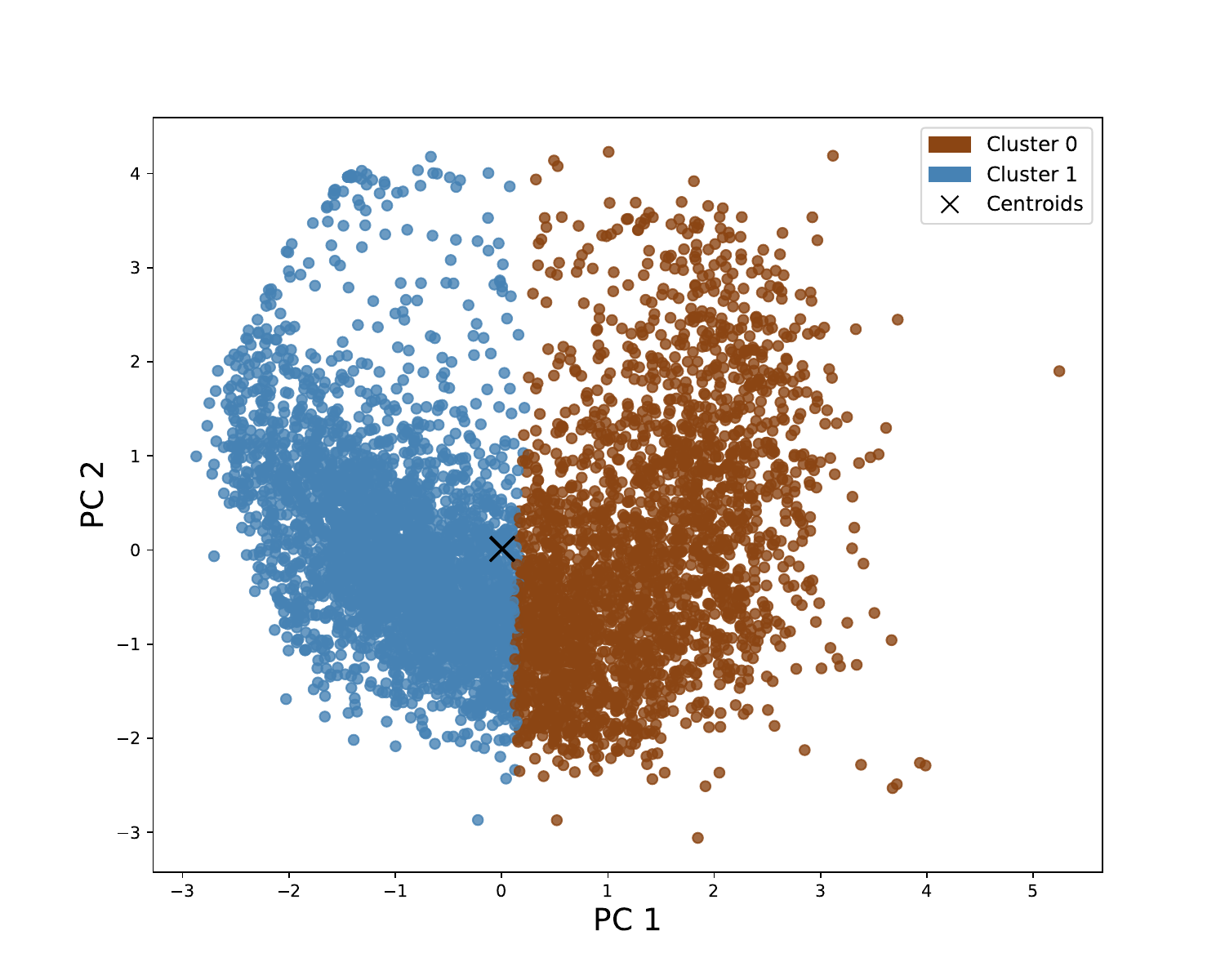}
        \caption{$k=2$}
    \end{subfigure}
    \begin{subfigure}{0.46\linewidth}
        \includegraphics[width=8cm]{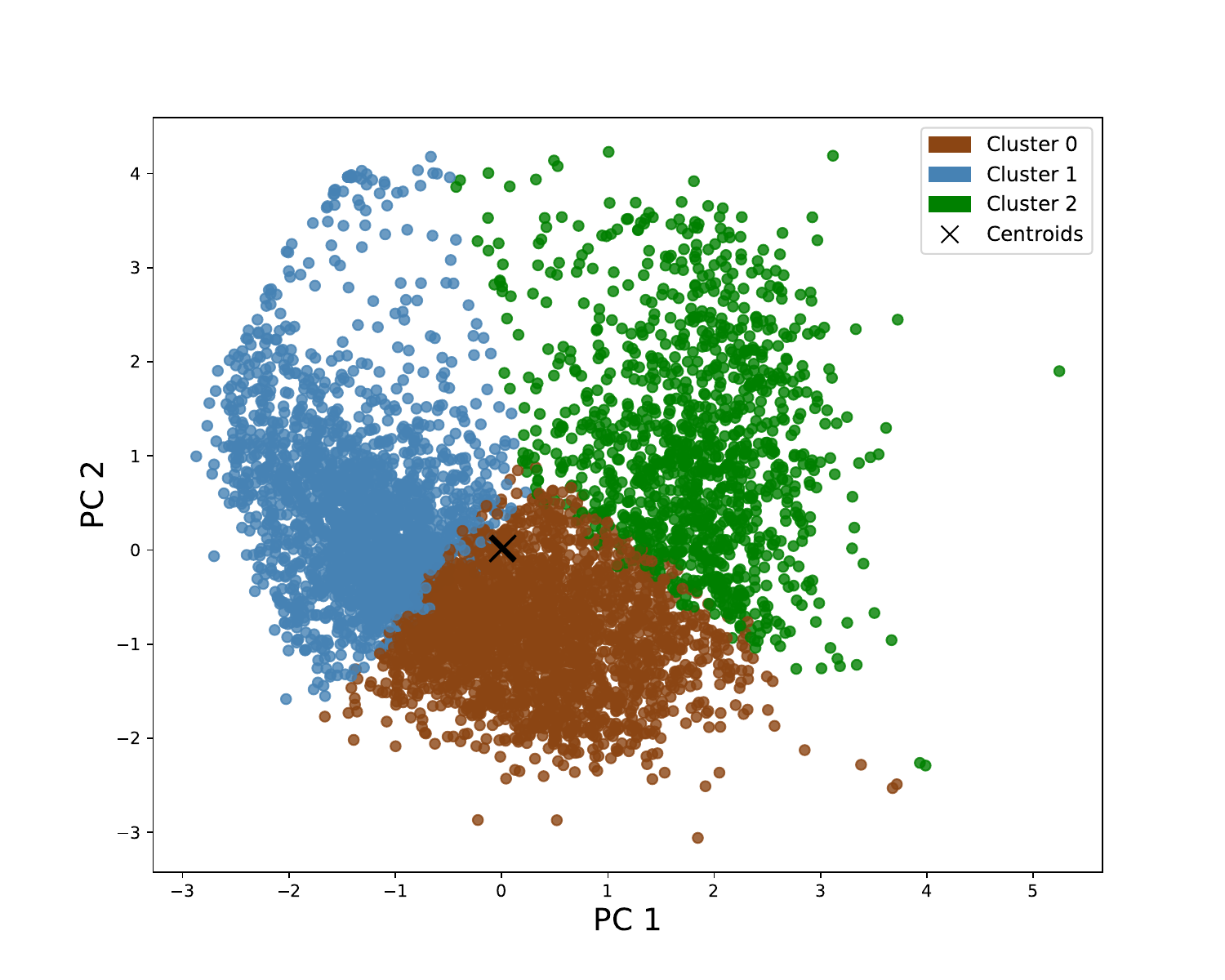}
        \caption{$k=3$}
    \end{subfigure}
    \caption{\textbf{PCA visualization} for projection of normalized price trajectories (in swap time) onto the principal components. For $k=2$ the two clusters correspond closely to the main distinction between \textit{Safe} and \textit{Honeypot} tokens.}
    \label{fig:pca}
\end{figure}
Figs \ref{fig:pca} show the projection of the time-series data onto the first two principal components for the cases of $k=2$ and $k=3$ clusters, respectively. 
In both cases, the clusters are well separated, confirming that the leading dimensions of variability in \textit{swap time} correspond closely to the clustering structure. 
The configuration with $k=2$ aligns with the primary distinction between \textit{Safe} and \textit{Honeypot} tokens, while $k=3$ captures additional intra-group heterogeneity. 
}

\begin{figure}[H]
    \centering
    \begin{subfigure}{0.46\linewidth}
        \includegraphics[width=8cm]{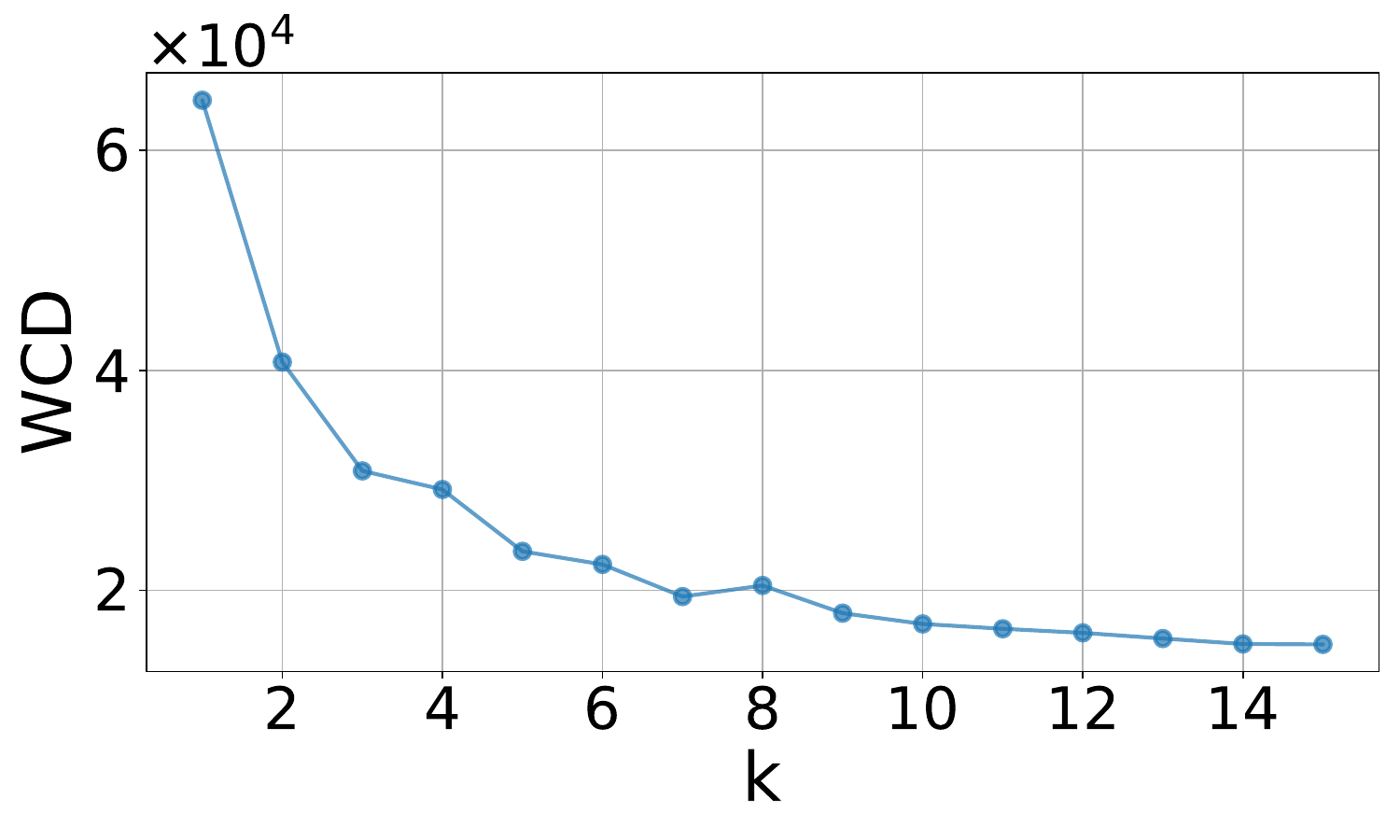}
        \caption{Honeypot tokens}
    \end{subfigure}
    \begin{subfigure}{0.46\linewidth}
        \includegraphics[width=8cm]{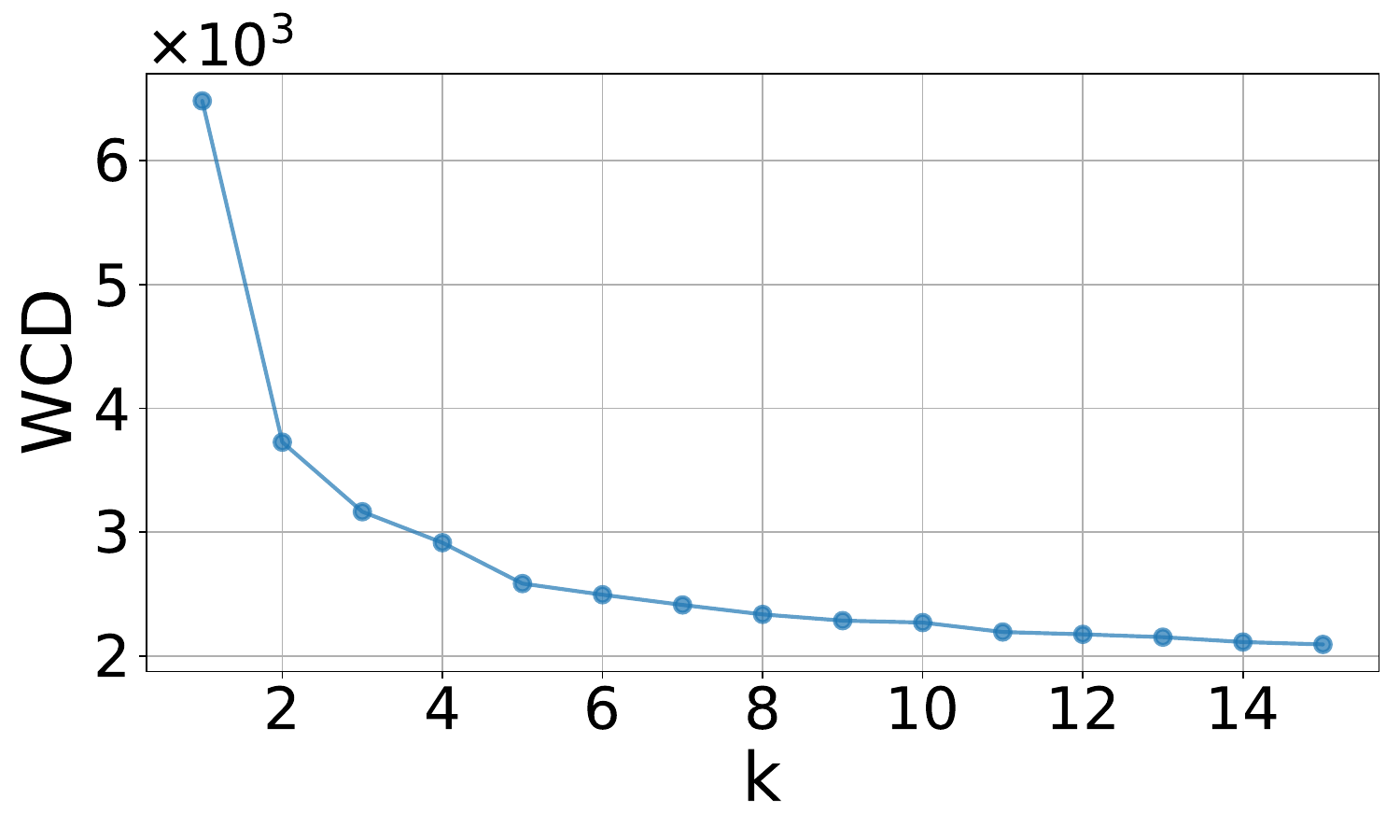}
        \caption{Sellable tokens}
    \end{subfigure}
    \caption{\textbf{Elbow curve.} Within-cluster Dissimilarity as a function of the number of clusters $k$. 
    The inflection point in $k=3$ indicates the region where additional clusters yield diminishing improvements in compactness.}
    \label{fig:DTW_WCD}
\end{figure}

{
The same elbow curve analysis in Fig. \ref{fig:DTW_WCD} is done for the DTW clustering method using a WCSS error metric with the Within-Cluster Dissimilarity metric:
\begin{align}
\label{eq:WCD}
WCD(k) = \sum_{c=1}^k \sum_{x_i \in \mathcal{C}_c} DTW(x_i, m_c)\,\,;
\end{align}
where $\mathcal{C}_c$ is the series set $\{x_i\}$ of the $c^{\rm th}$-cluster, $m_c$ the medoid (i.e. central series) of the $c^{\rm th}$-cluster and $DTW(x_i, m_c)$ is the DTW distance between $x_i$ and $m_c$.
}

\end{document}